\documentclass[aps,twocolumn,showkeys,showpacs,preprintnumbers,prd,superscriptaddress,nofootinbib,10pt]{revtex4-1}
\usepackage{graphicx,epsf,bm,amsmath,amsfonts,amssymb,epstopdf,natbib,hyperref,xcolor,appendix,verbatim, multirow,diagbox,booktabs,siunitx}
\usepackage[normalem]{ulem}
\usepackage{array}
\usepackage{diagbox}
\usepackage{makecell}
\usepackage[makeroom]{cancel}
\usepackage{fontawesome}
\usepackage{hyperref}
\hypersetup{colorlinks=true,urlcolor=magenta,citecolor=blue,linkcolor=blue,menucolor=blue,anchorcolor=blue,filecolor=blue}

\newcolumntype{?}{!{\vrule width 3pt}}

\usepackage{graphicx} 
\usepackage{array}    
\usepackage{makecell} 

\date{\today}

\def\@fnsymbol#1{{\ifcase#1\or \fmarki\or \fmarkii\or \fmarkiii\or \fmarkiv\or \fmarkv\or \else\@ctrerr\fi}}
\makeatother

\begin{document}

\title{Slow-Roll Inflation at Arbitrary Order}

\author{William Giarè}
\email{giare@hawaii.edu}
\affiliation{Department of Physics and Astronomy, University of Hawai‘i, Honolulu, HI 96822, USA \looseness=-1}

\begin{abstract}
\noindent 
We derive explicit expressions for the primordial scalar and tensor spectra at arbitrary finite order $p$ in the slow-roll expansion, valid both for canonical single-field inflation and $k$-inflation. At any order, the spectra are determined by known combinations of slow-roll parameters weighted by universal coefficients entirely fixed by the free de Sitter solutions, without requiring perturbatively corrected mode functions. Using these spectra, we obtain closed-form consistency relations that connect the tensor tilt $n_t$ and its runnings $\alpha_n^t\equiv d^n n_t/d(\ln k)^n$ to the tensor-to-scalar ratio $r$, the scalar spectral index $n_s$, and its runnings $\alpha_n^s\equiv d^n(n_s-1)/d(\ln k)^n$. The consistency relations are valid for arbitrary $n\geq 0$ and any finite $p\geq 0$, without recursion over lower-order relations. For $k$-inflation, they additionally depend on the scalar sound speed and its scale derivatives. We develop an algorithm that generates both the spectra and consistency relations for any pair $(n,p)$. We reproduce known results through next-to-next-to-next-to-leading order (N3LO) and extend the explicit calculations up to N10LO.
\\
\\
{\footnotesize Project Repository: \faGithub\ \url{https://github.com/williamgiare/slow-roll-inflation}}
\end{abstract}

\maketitle

\section{Introduction}
\label{sec:intro}

In canonical single-field slow-roll (SR) inflation~\cite{Kazanas:1980tx,Starobinsky:1980te,Sato:1981ds,Guth:1980zm,Mukhanov:1981xt,Linde:1981mu,Albrecht:1982wi}, the primordial scalar and tensor spectra at leading order (LO) read~\cite{Mukhanov:1981xt,Hawking:1982cz,Starobinsky:1982ee,Guth:1982ec,Bardeen:1983qw,Lidsey:1995np,Riotto:2002yw,Kinney:2009vz,Baumann:2009ds,Martin:2013tda,Senatore:2016aui}
\begin{equation}
\mathcal P_s(k)=\frac{H^2}{8\pi^2 M^2_{\rm Pl}\,\epsilon_1},
\qquad
\mathcal P_t(k)= \frac{2H^2}{\pi^2 M^2_{\rm Pl}},
\label{eq:LO_spectra}
\end{equation}
where $\epsilon_1\equiv-d\ln H/dN$, and $dN\equiv H\,dt$. Since $\epsilon_1$ is itself determined by the evolution of $H$, the scale dependence of both spectra is governed by the same quasi-de Sitter background.

This common background dependence gives rise to consistency relations between scalar and tensor spectral parameters~\cite{Riotto:2002yw,Cortes:2006ap,Kinney:2009vz,Baumann:2009ds,Martin:2013tda}. When evaluated at the same pivot scale $k_\star$, the tensor-to-scalar ratio $r\equiv\mathcal P_t/\mathcal P_s$ and the tensor tilt $n_t\equiv d\ln\mathcal P_t/d\ln k$ satisfy the standard LO consistency relation $r=-8n_t$. Successive derivatives with respect to $\ln k$ further generate an infinite hierarchy of relations connecting the tensor runnings $\alpha_n^t\equiv d^n n_t/d(\ln k)^n$ to the scalar spectral index $n_s$ and its runnings $\alpha_n^s\equiv d^n(n_s-1)/d(\ln k)^n$, all evaluated at the pivot $k_\star$.\footnote{In our notation, $\alpha_0^s=n_s(k_\star)-1$, $\alpha_1^s=\alpha_s(k_\star)$, $\alpha_2^s=\beta_s(k_\star)$, and so on, while $\alpha_0^t=n_t(k_\star)$, $\alpha_1^t=\alpha_t(k_\star)$, $\alpha_2^t=\beta_t(k_\star)$, and so on. At LO, $\alpha_0^t=-r/8$.} At LO, this hierarchy is generated by~\cite{Cortes:2006ap}
\begin{equation}
\frac{dn_t}{d\ln k}
=
n_t\left[n_t-(n_s-1)\right].
\label{eq:intro_flow}
\end{equation}

Corrections beyond LO have been extensively studied, broadly following two closely related directions. First, one can simply increase the SR order when deriving the primordial spectra. Higher-order expressions were obtained already in early works on inflationary perturbation theory~\cite{Stewart:1993bc} and subsequently used to derive higher-order consistency relations~\cite{Liddle:1994cr,Lidsey:1995np}. A systematic Green-function treatment was later developed in Ref.~\cite{Gong:2001he}, where the primordial spectra were organized in a form that can, in principle, be extended to arbitrary order in the SR expansion. A second path is to reconstruct the inflationary background from observables~\cite{Copeland:1993zn,Liddle:1994cr,Lidsey:1995np,Joy:2005ep,Joy:2005pe}. The Hubble-flow hierarchy introduced in Ref.~\cite{Kinney:2002qn} provides a systematic description of background evolution and was subsequently connected directly to observable spectral quantities in higher-order reconstruction methods~\cite{Peiris:2006ug,Peiris:2006sj}.

Despite these developments, some limitations remain. First, the Green-function approach requires perturbative corrections to the mode functions to be calculated order by order~\cite{Gong:2001he}. At present, explicit expressions for the scalar and tensor spectra have been obtained only through N3LO~\cite{Auclair:2022yxs}, with extensions to broader inflationary theories~\cite{Bianchi:2024qyp}. Second, the infinite hierarchy of consistency relations has been formulated only at LO and NLO~\cite{Cortes:2006ap}. At both orders, obtaining explicit expressions for higher tensor runnings from this hierarchy requires successive differentiation of the preceding relations.

In this work, we address both limitations. We first derive explicit expressions for the primordial scalar and tensor spectra at arbitrary finite order in the SR expansion. At each order, the spectra are determined by combinations of inflationary slow-roll parameters weighted by universal coefficients fixed entirely by the free de Sitter solutions, without requiring perturbatively corrected mode functions. We then show that a perturbative inversion allows the inflationary background quantities to be expressed directly in terms of spectral observables, yielding closed-form consistency relations between scalar and tensor spectral parameters at arbitrary derivative order and any finite order in the SR expansion. 

Denoting by $n\geq0$ the order in scale derivatives and by $p\geq0$ the order of SR corrections (with $p=0$ corresponding to LO), the highest absolute SR order is $W=n+1+p$, and the resulting relations take the form\footnote{The $n$th scale derivative starts at order $n+1$ in the SR expansion.}
\begin{equation}
\alpha_n^t
=
F_{n,p}
\left(
r,\alpha_0^s,\ldots,\alpha_{n-1+p}^s
\right)
+
\mathcal O\left(\mathrm{SR}^{n+p+2}\right).
\label{eq:intro_master}
\end{equation}

We further generalize our results to $k$-inflation with a slowly varying scalar sound speed, showing that the universal coefficients in the primordial spectra remain unchanged, while the consistency relations additionally depend on the sound speed and its scale evolution. 

We develop an algorithm that generates both the spectra and the corresponding consistency relations at any specified $(n,p)$, without recursion over lower-order relations. For both the canonical case and $k$-inflation, we reproduce known results through N3LO while extending them to arbitrary order.

\section{Closed-Form Consistency Relations at Leading Order}
\label{sec:leading_order}

To illustrate the main idea in its simplest form, we begin with the LO case (i.e., $p=0$). This case is somewhat special because the dependence on the inflationary background, namely on $H$ and its derivatives, can be eliminated before constructing the hierarchy. Starting from the LO spectra in Eq.~\eqref{eq:LO_spectra}, the consistency condition can be written directly in terms of observable spectral quantities as $n_t=-r/8
=-\mathcal P_t / (8\mathcal P_s)$, which implies
\begin{equation}
\frac{d}{d\ln k}
\left(
\frac{1}{\mathcal P_t}
\right)
=
\frac{1}{8\mathcal P_s}.
\label{eq:inverse_tensor_flow}
\end{equation}
Defining $L\equiv\ln(k/k_\star)$ and integrating directly between $k_\star$ and $k$, we get~\cite{Copeland:1993ie,Copeland:1993jj,Lidsey:1995np,Brooker:2016imi}
\begin{equation}
\frac{\mathcal P_t(k_\star e^L)}
{\mathcal P_t(k_\star)}
=
\left[
1-\alpha_0^t
\int_0^L
\frac{\mathcal P_s(k_\star)}
{\mathcal P_s(k_\star e^u)}
du
\right]^{-1}.
\label{eq:finite_consistency}
\end{equation}
Equation~\eqref{eq:finite_consistency} expresses the LO consistency condition as a finite transport relation between two scales and provides a direct route to the complete derivative hierarchy.

Assuming analyticity in a neighborhood of the pivot, the scalar and tensor spectra can be expanded as
\begin{equation}
\ln
\frac{\mathcal P_A(k_\star e^L)}
{\mathcal P_A(k_\star)}
=
\sum_{n=0}^{\infty}
\alpha_n^A
\frac{L^{n+1}}{(n+1)!},
\label{eq:spectral_expansion}
\end{equation}
where $A=s,t$ labels the scalar and tensor spectra, respectively.

To expand the scalar factor entering Eq.~\eqref{eq:finite_consistency}, we define $K_0\equiv1$ and $K_n \equiv B_n
\left(
-\alpha_0^s,
-\alpha_1^s,
\ldots,
-\alpha_{n-1}^s
\right)$ for $n\geq1$. Here $B_n$ denotes the complete exponential Bell polynomial.\footnote{The complete exponential Bell polynomial are defined by
\[
\exp\left[
\sum_{j\geq1}
\frac{x_j t^j}{j!}
\right]
=
\sum_{n\geq0}
\frac{t^n}{n!}
B_n(x_1,\ldots,x_n).
\]
}
Equation~\eqref{eq:spectral_expansion} then gives
\begin{equation}
\frac{\mathcal P_s(k_\star)}
{\mathcal P_s(k_\star e^L)}
=
\sum_{n=0}^{\infty}
K_n
\frac{L^n}{n!},
\label{eq:scalar_inverse_expansion}
\end{equation}
and therefore
\begin{equation}
\int_0^L
\frac{\mathcal P_s(k_\star)}
{\mathcal P_s(k_\star e^u)}
du
=
\sum_{n=0}^{\infty}
K_n
\frac{L^{n+1}}{(n+1)!}.
\label{eq:integrated_scalar_expansion}
\end{equation}

Taking the logarithm of Eq.~\eqref{eq:finite_consistency} and using Eqs.~\eqref{eq:spectral_expansion} and~\eqref{eq:integrated_scalar_expansion}, we obtain
\begin{equation}
\sum_{n=0}^{\infty}
\alpha_n^t
\frac{L^{n+1}}{(n+1)!}
=
-\ln
\left[
1-\alpha_0^t
\sum_{n=0}^{\infty}
K_n
\frac{L^{n+1}}{(n+1)!}
\right].
\label{eq:generating_relation}
\end{equation}

Using $-\ln(1-x)=\sum_{q\geq1}x^q/q$ and the partial exponential Bell polynomials $B_{m,q}$,\footnote{The partial exponential Bell polynomials are defined by
\[
\left(
\sum_{j\geq1}
\frac{t^j}{j!}
x_j
\right)^q
=
q!
\sum_{m\geq q}
\frac{t^m}{m!}
B_{m,q}(x_1,\ldots,x_{m-q+1}).
\]
}
with $x_j=K_{j-1}$, comparison of equal powers of $L$, together with $\alpha_0^t=-r/8$, gives
\begin{equation}
F_{n,0}
\equiv 
\sum_{q=1}^{n+1}
\left(
-\frac{r}{8}
\right)^q
(q-1)! \,
B_{n+1,q}
\left(
K_0,\ldots,K_{n+1-q}
\right),
\label{eq:LO_closed_form}
\end{equation}
which gives the $p=0$ case of Eq.~\eqref{eq:intro_master}.

Equation~\eqref{eq:LO_closed_form} provides a closed representation of the recursive LO consistency hierarchy generated by Eq.~\eqref{eq:intro_flow}, valid at arbitrary $n$ without successively constructing lower-order tensor runnings. While the recursive hierarchy is well known~\cite{Cortes:2006ap}, to the best of our knowledge this explicit closed form has not been given previously.

The LO derivation is particularly transparent because the finite scale-transport relation~\eqref{eq:finite_consistency} eliminates the inflationary background before constructing the hierarchy, which then follows directly from its Taylor expansion. Beyond LO, however, corrections to $\alpha_0^t=-r/8$ introduce additional dependence on the background evolution, and Eq.~\eqref{eq:inverse_tensor_flow} no longer closes directly in terms of the spectra. Instead, the spectra must first be computed at the required SR order, and the background dependence subsequently eliminated in favor of observables. In what follows, we develop this general construction for arbitrary $(n,p)$.

\section{Spectra and Consistency Relations at Arbitrary Order}
\label{sec:arbitrary_order}

Beyond LO, the spectra acquire corrections involving higher derivatives of the inflationary background. To organize these contributions, we introduce the Hubble-flow hierarchy~\cite{Schwarz:2001vv,Kinney:2002qn},
\begin{equation}
\epsilon_{i+1}
\equiv
\frac{d\ln|\epsilon_i|}{dN},
\qquad i\geq1.
\label{eq:hubble_flow}
\end{equation}
Since $d\epsilon_i/dN=\epsilon_i\epsilon_{i+1}$, the derivative with respect to $N$ of any function $X$ of the Hubble-flow parameters can be written as
\begin{equation}
D_N X
\equiv
\frac{dX}{dN}
=
\sum_{i\geq1}
\epsilon_i\epsilon_{i+1}
\frac{\partial X}{\partial\epsilon_i}.
\label{eq:DN}
\end{equation}
Counting each $\epsilon_i$ as first order in SR, differentiation with respect to $N$ raises the SR order by one. More generally, if $X$ starts at order $m$, then $D_N^j X=\mathcal O\left(\mathrm{SR}^{m+j}\right)$.

This counting makes the arbitrary-order construction finite. Although the Hubble-flow hierarchy contains infinitely many parameters, only a finite number can contribute at any fixed $(n,p)$.

Using this counting, we generalize the LO result to arbitrary finite SR order $p$. We first derive explicit expressions for the primordial scalar and tensor spectra in terms of the inflationary background parameters at arbitrary SR order. Then we eliminate their background dependence in favor of observables and derive the consistency relation $F_{n,p}$ for arbitrary $n$ and finite $p$.

\subsection{Primordial Spectra at Arbitrary Order}
\label{sec:Spectra}

We first show that the primordial scalar and tensor spectra can be expressed explicitly in terms of a finite set of Hubble-flow parameters at any fixed SR order $W$.

We start from the Mukhanov--Sasaki equations for scalar and tensor modes~\cite{Sasaki:1986hm,Mukhanov:1988jd}
\begin{equation}
v_s''+
\left(
k^2-\frac{z_s''}{z_s}
\right)v_s
=
0, \qquad v_t''+
\left(
k^2-\frac{a''}{a}
\right)v_t
=
0.
\end{equation}
Here primes denote derivatives with respect to conformal time $\eta$, and we define $z_s\equiv a\sqrt{2\epsilon_1}M_{\rm Pl}$. All dependence on the inflationary background is encoded in the effective potentials, $z_s''/z_s$ for scalar modes and $a''/a$ for tensor modes. Introducing $x\equiv-k\eta$ and the dimensionless mode functions $y_A\equiv\sqrt{2k}\,v_A$, with $A=s,t$, the scalar and tensor equations can be written as
\begin{equation}
\frac{d^2y_A}{dx^2}
+
\left(
1-\frac{2}{x^2}
\right)y_A
=
\frac{g_A(x)}{x^2}y_A.
\label{eq:MS_source}
\end{equation}
The left-hand side is the exact de Sitter equation, while all SR corrections are contained in the source $g_A$. Defining $h\equiv-\eta aH$, the scalar source is
\begin{equation}
g_s
=
h^2
\left[
2-\epsilon_1
+\frac{3-\epsilon_1}{2}\epsilon_2
+\frac{\epsilon_2^2}{4}
+\frac{\epsilon_2\epsilon_3}{2}
\right]
-2,
\label{eq:gs}
\end{equation}
while the tensor source reads
\begin{equation}
g_t
=
h^2(2-\epsilon_1)-2.
\label{eq:gt}
\end{equation}
The quantity $h$ obeys $D_Nh=h(1-\epsilon_1)-1$ and is therefore determined order by order by the same Hubble-flow hierarchy, with $h=1+\mathcal O(\mathrm{SR})$.\footnote{In the exact de Sitter limit, $\epsilon_i=0$ and $h=1$, so that both sources vanish.} In Appendix~\ref{app:spectra_proof}, we show that $h$ admits an explicit finite representation in terms of complete Bell polynomials at any fixed SR order.

Equation~\eqref{eq:MS_source} can be solved perturbatively around the de Sitter solution with Bunch--Davies initial conditions~\cite{Gong:2001he}. The primordial spectra are determined by the constant super-horizon modes. For scalar perturbations, the quantity that freezes is $v_s/z_s$, while for tensor perturbations it is $v_t/a$. It is therefore convenient to normalize the corresponding pump fields as $Z_s(x)\equiv z_s(x)/z_s(1)$ and $Z_t(x)\equiv a(x)/a(1)$, and define~\cite{Habib:2004hd}
\begin{equation}
J_A
\equiv
-i\lim_{x\rightarrow0}
\frac{y_A(x)}{Z_A(x)}.
\label{eq:J}
\end{equation}
$J_A$ isolates the constant super-horizon amplitude that directly determines the corresponding primordial spectrum~\cite{Habib:2004hd}. The scalar and tensor spectra can then be written as~\cite{Stewart:1993bc,Auclair:2022yxs}
\begin{equation}
\mathcal P_s
=
\frac{H^2}
{8\pi^2M_{\rm Pl}^2\epsilon_1}
e^{M_s},
\qquad
\mathcal P_t
=
\frac{2H^2}
{\pi^2M_{\rm Pl}^2}
e^{M_t},
\label{eq:spectra_general}
\end{equation}
where $M_A\equiv\ln|J_A|^2-2\ln h$, and all background quantities are evaluated at the reference point $x=1$.

To determine $M_A$ at arbitrary SR order, we expand the source for each mode $k$ around the reference point $x=1$.\footnote{The choice $x=1$ is only a reference point for the SR expansion. Since $k=aH/h$ at $x=1$, it coincides with exact horizon crossing only in the de Sitter limit.} The expansion reads
\begin{equation}
g_A(x)
=
\sum_{j=0}^{W-1}
g_{A,j}\,(\ln x)^j
+
\mathcal O\left(\mathrm{SR}^{W+1}\right),
\label{eq:g_expansion}
\end{equation}
where $g_{A,j}\equiv \frac{1}{j!}[(-hD_N)^j g_A]_{\leq W}$. Here $[\cdots]_{\leq W}$ means that, after expanding in SR, only terms up to absolute order $W$ are retained. Since $g_A$ starts at first order in SR and $d/d\ln x=-hD_N$, each coefficient $g_{A,j}$ starts at order $j+1$. Consequently, only finitely many source factors contribute to the super-horizon amplitude at any fixed order $W$, and $M_A$ can be written as
\begin{equation}
[M_A]_{\leq W}
=
\left[
-2\ln h
+
\sum_{m=1}^{W} \,
\sum_{|\mathbf j|\leq W-m} \,
\chi_{m;\mathbf j} \,
\prod_{i=1}^{m}g_{A,j_i}
\right]_{\leq W},
\label{eq:MA_universal_main}
\end{equation}
where $\mathbf j=(j_1,\ldots,j_m)$, with $j_i\geq 0$, and $|\mathbf j|=\sum_i j_i$. Here $m$ counts the number of source factors, while the coefficients $\chi_{m;\mathbf j}$ are universal and depend only on the free de Sitter solutions. All dependence on the inflationary background is contained in $h$ and $g_{A,j}$. In Appendix~\ref{app:spectra_proof}, we derive these coefficients explicitly at arbitrary order, without requiring corrected mode functions or lower-order amplitudes.

Once the primordial spectra are known in terms of the Hubble-flow parameters, all spectral observables follow by differentiation with respect to $\ln k$. At the reference point $x=1$, differentiation with respect to scale is related to the background evolution by $d/d\ln k=hD_N$. Using Eq.~\eqref{eq:spectra_general}, the tensor-to-scalar ratio is
\begin{equation}
r
=
16\epsilon_1 e^{M_t-M_s},
\label{eq:r_background}
\end{equation}
while the scalar and tensor spectral indices are
\begin{equation}
\alpha_0^s
=
h
\left(
-2\epsilon_1-\epsilon_2+D_NM_s 
\right),
\label{eq:ns_background}
\end{equation}
and
\begin{equation}
\alpha_0^t
=
h
\left(
-2\epsilon_1+D_NM_t
\right).
\label{eq:nt_background}
\end{equation}
Higher runnings then follow as
\begin{equation}
\alpha_j^s
=
(hD_N)^j\,\alpha_0^s,
\qquad
\alpha_j^t
=(hD_N)^j\,\alpha_0^t.
\label{eq:spectral_background}
\end{equation}

Since $M_s$ and $M_t$ are known perturbatively at order $W$, these relations determine the spectral observables explicitly in terms of a finite subset of the Hubble-flow parameters. The construction therefore defines the forward map
\begin{equation}
\{\epsilon_i\}
\longrightarrow
\{\mathcal P_s,\mathcal P_t\}
\longrightarrow
\{r,\alpha_0^s,\alpha_1^s,\ldots,\alpha_0^t,\alpha_1^t,\ldots\}
\end{equation}
at any order $W$. Denoting the background variables (i.e., the Hubble-flow parameters required at that order) by $\mathbf b$, and the spectral observables by $\mathbf o$, we write this map as
\begin{equation}
\mathbf o
=
\mathcal U_W(\mathbf b).
\label{eq:UW}
\end{equation}

\subsection{Eliminating the Inflationary Background}
\label{sec:Background_removal}

Our goal now is to construct the inverse map,
\begin{equation}
\mathbf b
=
\mathcal U_W^{-1}(\mathbf o),
\label{eq:UW_inverse}
\end{equation}
and express any quantity that depends on the inflationary background directly in terms of the spectral observables. For the inversion, we focus on the specific observable vector $\mathbf o
=
\left(
r,\alpha_0^s,\ldots,\alpha_{W-1}^s
\right)$.

For the background variables, instead of using the Hubble-flow parameters $\epsilon_i$, it is convenient to use $\epsilon_1$ and the successive derivatives of $\epsilon_2$ defined as
\begin{equation}
q_j
\equiv
D_N^{j-1}\epsilon_2,
\qquad
j\geq1.
\label{eq:q_definition}
\end{equation}
By construction, $q_1=\epsilon_2$ and $q_j=\mathcal O(\mathrm{SR}^j)$. In these coordinates our background vector therefore reads $\mathbf b=\left(\epsilon_1,q_1,\ldots,q_W\right)$.

We proceed in two steps. We first explicitly construct the inverse map at LO. We then use this LO inverse map as the starting point for a perturbative construction of $\mathcal U_W^{-1}$ at arbitrary order $W$.

\subsubsection{Leading-Order Inversion}

At LO, the first background variable follows directly from $r=16\epsilon_1$, giving
$\epsilon_1=r/16$. Moreover, from $\alpha_0^s=-2\epsilon_1-\epsilon_2$ and $\alpha_0^t=-2\epsilon_1$ we get $\epsilon_2=\alpha_0^t-\alpha_0^s$. Since at LO $d/d\ln k=D_N$, Eq.~\eqref{eq:q_definition} implies $q_j = \alpha_{j-1}^t-\alpha_{j-1}^s$ for $j\geq1$. Using the closed LO consistency relation~\eqref{eq:LO_closed_form} we obtain
\begin{equation}
q_j
=
F_{j-1,0}
\left(
r,\alpha_0^s,\ldots,\alpha_{j-2}^s
\right)
-\alpha_{j-1}^s,
\qquad j\geq1.
\label{eq:q_LO_closed}
\end{equation}
For $j=1$, the scalar arguments of $F_{0,0}$ are absent.
Together with $\epsilon_1=r/16$, Eq.~\eqref{eq:q_LO_closed} determines every component of $\mathbf b$ directly in terms of $\mathbf o$ and therefore defines the explicit LO inverse map
\begin{equation}
\mathcal U_{\rm LO}^{-1}:
\mathbf o
\longrightarrow
\mathbf b.
\label{eq:ULO_inverse}
\end{equation}

\subsubsection{Arbitrary-Order Inversion}

We now use the LO inverse to construct the inverse at arbitrary order $W$.

Consider a background $\mathbf b$. The full forward map~\eqref{eq:UW} determines the corresponding observables $\mathbf o$. These observables are known. We can apply to them the LO inverse~\eqref{eq:ULO_inverse}, which is also known. We denote the resulting reconstructed background by
\begin{equation}
\widetilde{\mathbf b}
\equiv
\mathcal U_{\rm LO}^{-1}(\mathbf o).
\label{eq:b_tilde}
\end{equation}

Using Eq.~\eqref{eq:UW}, the relation between the original and reconstructed backgrounds is therefore $\widetilde{\mathbf b}
=
\mathcal U_{\rm LO}^{-1}
\!\left[
\mathcal U_W(\mathbf b)
\right]$.
Since $\mathcal U_{\rm LO}^{-1}$ exactly inverts the LO part of $\mathcal U_W$, the reconstructed background differs from the original one only by higher-order SR corrections,
\begin{equation}
\widetilde{\mathbf b}
=
\mathbf b+\delta\mathbf b_W(\mathbf b),
\label{eq:background_shift}
\end{equation}
where each component of $\delta\mathbf b_W$ starts at least one SR order above the corresponding component of $\mathbf b$.

Let $f(\mathbf b)$ be any function admitting a regular SR expansion. The shift in Eq.~\eqref{eq:background_shift} induces a corresponding change in $f$, which we define as
\begin{equation}
\Delta_W f(\mathbf b)
\equiv
f\!\left[
\mathbf b+\delta\mathbf b_W(\mathbf b)
\right]
-
f(\mathbf b).
\label{eq:DeltaW}
\end{equation}
Equivalently, we can write
\begin{equation}
f(\widetilde{\mathbf b})
=
(I+\Delta_W)f(\mathbf b).
\label{eq:Delta_action}
\end{equation}

Since $\delta\mathbf b_W$ is higher order in SR, each application of $\Delta_W$ raises the SR order by at least one. If $f$ starts at order $d$, then $\Delta_W f,\Delta_W^2 f,\ldots,\Delta_W^\ell f$ start at least at orders $d+1,d+2,\ldots,d+\ell$, respectively. The inverse of $I+\Delta_W$ is therefore given, at order $W$, by the finite geometric expansion
\begin{equation}
(I+\Delta_W)^{-1}f
=
\left[
\sum_{\ell=0}^{W-d}
(-\Delta_W)^\ell f
\right]_{\leq W}.
\label{eq:finite_inverse}
\end{equation}
The sum terminates because all subsequent terms lie beyond order $W$.

Using Eq.~\eqref{eq:finite_inverse}, we can express the action of the full inverse map on any function $f$ of the inflationary background, starting at SR order $d$. We therefore define the background-elimination map
\begin{equation}
E_W[f](\mathbf o)
\equiv
\left(
\left[
\sum_{\ell=0}^{W-d}
(-\Delta_W)^\ell f
\right]_{\leq W}
\right)
\left[
\mathcal U_{\rm LO}^{-1}(\mathbf o)
\right].
\label{eq:EW}
\end{equation}
By construction, $E_W[f](\mathbf o)=f[\mathcal U_W^{-1}(\mathbf o)]+\mathcal O(\mathrm{SR}^{W+1})$, expressing the result directly in terms of the observable vector $\mathbf o$.

A detailed derivation of Eq.~\eqref{eq:finite_inverse}, together with an explicit finite representation of $E_W[f]$ in terms of ordinary derivatives and the known shift $\delta\mathbf b_W$, is provided in Appendix~\ref{app:Consistency_Relations}.

\subsection{Consistency Relations at Arbitrary Order}
\label{sec:Relations}

We now have all the ingredients required to derive the consistency relations at arbitrary order. From the tensor spectrum, the $n$th tensor spectral running can be written in terms of the inflationary background as
\begin{equation}
\alpha_n^t
=
(hD_N)^n
\left[
h
\left(
-2\epsilon_1+D_NM_t
\right)
\right].
\label{eq:tensor_hierarchy_background}
\end{equation}
Equation~\eqref{eq:tensor_hierarchy_background} does not require lower-order tensor consistency relations as input. However, it still involves $n$ successive applications of the scale-derivative operator $hD_N$.

These successive derivatives can be evaluated directly. For any function $X$ of the inflationary background and any integer $\nu\geq1$,
\begin{equation}
(hD_N)^\nu X
=
\sum_{k=1}^{\nu}
c_{\nu,k}D_N^kX,
\label{eq:closed_derivative}
\end{equation}
where the coefficients $c_{\nu,k}$ are explicit finite functions of the background. Their closed expressions and derivation are given in Appendix~\ref{app:Consistency_Relations}.

Using $D_N\ln H=-\epsilon_1$, Eq.~\eqref{eq:tensor_hierarchy_background} can be written as $(hD_N)^{n+1}(2\ln H+M_t)$. Applying Eq.~\eqref{eq:closed_derivative} with $\nu=n+1$ therefore gives
\begin{equation}
\alpha_n^t
=
\sum_{k=1}^{n+1}
c_{n+1,k}
\left(
-2D_N^{k-1}\epsilon_1
+
D_N^kM_t
\right).
\label{eq:tensor_closed}
\end{equation}
Equation~\eqref{eq:tensor_closed} is a finite expression valid for arbitrary $n$. The successive applications of $hD_N$ have been replaced by ordinary background derivatives weighted by explicitly known coefficients. At fixed SR order, these derivatives admit explicit finite expressions, as detailed in Appendix~\ref{app:Consistency_Relations}.

The right-hand side of Eq.~\eqref{eq:tensor_closed} is now an explicit function of the inflationary background. Applying the finite inversion $E_W$, with $W=n+1+p$, we obtain
\begin{align}
\nonumber & F_{n,p}
\left(
r,\alpha_0^s,\ldots,\alpha_{n+p-1}^s
\right)
\equiv\\&
E_W
\left[
\sum_{k=1}^{n+1}
c_{n+1,k}
\left(
-2D_N^{k-1}\epsilon_1
+
D_N^kM_t
\right)
\right]
(\mathbf o).
\label{eq:Fnp_definition}
\end{align}
Equation~\eqref{eq:Fnp_definition} provides the explicit function entering the general consistency relation~\eqref{eq:intro_master}, valid for arbitrary $n$ and any finite $p$. All sums are finite, and no lower-order tensor consistency relations are required.

The formalism developed in this section extends directly to single-field $k$-inflation with a slowly varying scalar sound speed. In acoustic time, the scalar mode equation maintains the same free de Sitter operator and universal coefficients, while the consistency relations additionally require the sound-speed hierarchy. The generalization is detailed in Appendix~\ref{app:noncanonical}.

\section{Implementation and Conclusions}
\label{sec:Implementation}

We implement the formalism in a symbolic algorithm for both canonical inflation and $k$-inflation. For any specified pair $(n,p)$, the algorithm generates the primordial spectra and the corresponding spectral observables, maintaining all contributions through the required SR order, and derives the consistency relations without recursion over lower-order relations. We explicitly reproduce known results up to N3LO and extend the calculation to N10LO. The symbolic code, validation tests, and corresponding spectra and consistency relations are publicly available in the project repository. Details of the validation are provided in Appendix~\ref{app:validation}.

Our results show that single-field slow-roll perturbation theory can be solved explicitly for two-point correlation functions at any finite order, without constructing lower orders first. The primordial scalar and tensor spectra are expressed through universal, model-independent coefficients fixed by the free de Sitter solutions, both in canonical inflation and in $k$-inflation. A finite perturbative inversion then yields closed-form consistency relations at arbitrary order in both slow-roll corrections and scale derivatives.

\begin{acknowledgments}
\noindent I thank Jeremy Sakstein and Sunny Vagnozzi for valuable suggestions and feedback on an early version of the project. I acknowledge support from National Aeronautics and Space Administration (NASA) under Grant No.~80NSSC24K0898.

The research problem, central ideas, and overall implementation strategy presented in this paper were developed by the author. OpenAI ChatGPT and Codex (GPT-6 Astra and GPT-6.1 Sol), together with Anthropic Claude (Opus 5.5), were used to cross-check derivations, clarify intermediate mathematical steps, and assist with code development and debugging. The author takes full responsibility for the results, the scientific conclusions, and the final content of the manuscript.
\end{acknowledgments}
\newpage
\onecolumngrid
\makeatletter
\def\set@footnotewidth{\onecolumngrid}
\makeatother
\appendix

\section{Derivation of the Arbitrary-Order Primordial Spectra}
\label{app:spectra_proof}

In this appendix, we derive the explicit finite expressions underlying the primordial spectra presented in Sec.~\ref{sec:Spectra} at arbitrary fixed SR order $W$. We begin by recalling the equations governing scalar and tensor perturbations in canonical single-field inflation. The corresponding Mukhanov--Sasaki variables $v_{A,k}$, with $A=s,t$, satisfy
\begin{equation}
\frac{d^2v_{A,k}}{d\eta^2}+\left(k^2-\frac{1}{z_A}\frac{d^2z_A}{d\eta^2}\right)v_{A,k}=0,
\end{equation}
where $\eta$ is conformal time, $k$ is the comoving wavenumber, and $z_A$ denotes the corresponding background pump field. For scalar perturbations, $z_s=aM_{\rm Pl}\sqrt{2\epsilon_1}$, where $a$ is the scale factor and $\epsilon_1=-\dot H/H^2$ is the first Hubble-flow parameter, while for tensor perturbations $z_t\propto a$.

To separate the free de Sitter evolution from the corrections generated by the inflationary background, we introduce the dimensionless variable $x=-k\eta$ and the rescaled mode function $y_A(x)=\sqrt{2k}\,v_{A,k}(\eta)$. We also define the normalized pump $Z_A(x)=z_A(\eta)/z_A(\eta_1)$, where $\eta_1=-1/k$, so that $Z_A(1)=1$. In terms of these variables, the mode and pump equations become
\begin{equation}
y_A''+\left(1-\frac{2}{x^2}\right)y_A=V_Ay_A,\qquad Z_A''-\frac{2}{x^2}Z_A=V_AZ_A,
\label{eq:app_mode_pump_equations}
\end{equation}
where primes now denote derivatives with respect to $x$, and $V_A(x)=g_A(x)/x^2$, with $g_A(x)=x^2Z_A''/Z_A-2$. The source $g_A$ therefore contains the corrections to the free de Sitter equations arising from the inflationary background.

In the free de Sitter limit, $g_A=0$, and both equations admit exact solutions. The mode function satisfying the Bunch--Davies condition $y_A(x)\to e^{ix}$ as $x\to\infty$ is
\begin{equation}
y_0(x)=\left(1+\frac{i}{x}\right)e^{ix},
\label{eq:app_BD_condition}
\end{equation}
while the corresponding growing pump solution, normalized at $x=1$, is $Z_0(x)=1/x$. These free solutions provide the starting point for our arbitrary-order construction as they will determine the Green functions used to calculate the corrections induced by $g_A$.

Solving the mode equations with the Bunch--Davies initial conditions and evaluating the resulting perturbations in the super-horizon limit, the scalar and tensor primordial spectra can be written as
\begin{equation}
\mathcal P_s(k)=\left.\frac{H^2}{8\pi^2M_{\rm Pl}^2\epsilon_1}\right|_{x=1}e^{M_s(k)},\qquad
\mathcal P_t(k)=\left.\frac{2H^2}{\pi^2M_{\rm Pl}^2}\right|_{x=1}e^{M_t(k)}.
\end{equation}
Here the corrections $M_A$, with $A=s,t$, are determined by the super-horizon amplitudes through
\begin{equation}
M_A=-2\ln h+\ln|J_A|^2,
\end{equation}
where $h=-\eta aH$ and
\begin{equation}
J_A=-i\lim_{x\to0}\frac{y_A(x)}{Z_A(x)}.
\label{eq:app_J}
\end{equation}
The factor $h$ accounts for the relation between conformal time and the background evolution, whereas $J_A$ determines the corrections to the super-horizon mode amplitude.

Our goal is to determine $M_A$ explicitly at arbitrary fixed SR order $W$. This requires two ingredients. First, in Sec.~\ref{sec:app_background_dependence} we derive the SR expansion of $h$ and the source coefficients $g_{A,j}$, which contain the dependence on the inflationary background. Second, in Sec.~\ref{sec:universal_coefficients} we calculate the universal coefficients $\chi_{m;\mathbf j}$ that determine $\ln|J_A|^2$ directly from the free de Sitter solutions. Combining these results provides the primordial spectra at any fixed SR order without requiring the iterative calculation of perturbatively corrected mode functions.

\subsection{The background dependence}
\label{sec:app_background_dependence}
The first ingredient is the background quantity $h=-\eta aH$, which determines the relation between conformal time and the evolution of the Hubble-flow parameters. Its SR expansion follows from $D_Nh=h(1-\epsilon_1)-1$ and can be written as
\begin{equation}
[h]_{\leq W}
=
1+\sum_{\ell=1}^{W}
B_\ell\!\left(
\epsilon_1,D_N\epsilon_1,\ldots,D_N^{\ell-1}\epsilon_1
\right).
\label{eq:app_h_closed}
\end{equation}
Expanding Eq.~\eqref{eq:app_h_closed} gives $h=1+\epsilon_1+\epsilon_1^2+\epsilon_1\epsilon_2+\cdots$. To verify this result, we use the Bell-polynomial identity $(D_N+\epsilon_1)B_\ell=B_{\ell+1}$. Acting with $D_N+\epsilon_1$ on the sum generates each successive term in the expansion, showing that Eq.~\eqref{eq:app_h_closed} satisfies $h=1+(D_N+\epsilon_1)h$ up to corrections of order $\mathcal O(\mathrm{SR}^{W+1})$.

The second ingredient is the logarithmic expansion of the scalar and tensor sources around the reference point $x=1$, introduced in Eq.~\eqref{eq:g_expansion}. At any fixed SR order $W$, this expansion takes the form
\begin{equation}
g_A(x)=\sum_{j=0}^{W-1}g_{A,j}(\ln x)^j+\mathcal O(\mathrm{SR}^{W+1}).
\end{equation}
The coefficients $g_{A,j}$ are obtained by differentiating the sources with respect to $\ln x$. Using $d/d\ln x=-hD_N$, we find
\begin{equation}
g_{A,j}\equiv\frac{1}{j!}\left[(-hD_N)^j g_A\right]_{\leq W},
\end{equation}
where all background quantities are evaluated at $x=1$. Since $g_A$ starts at first order in SR and each successive derivative raises the SR order by one, the coefficient $g_{A,j}$ starts at order $j+1$. Consequently, only terms with $j\leq W-1$ contribute at order $W$, making the expansion finite.

Having determined the SR expansion of $h$ and the source coefficients $g_{A,j}$, we can now combine these ingredients to obtain the corrections $M_A$ entering the primordial spectra in Eq.~\eqref{eq:spectra_general}. These corrections consist of the normalization term $-2\ln h$ and a sum of products of source coefficients, each weighted by a coefficient $\chi_{m;\mathbf j}$. At any fixed SR order $W$, the result takes the finite form
\begin{equation}
[M_A]_{\leq W}
=
\left[
-2\ln h+
\sum_{m=1}^{W}
\sum_{\substack{\mathbf j\in\mathbb N_0^m\\|\mathbf j|\leq W-m}}
\chi_{m;\mathbf j}
\prod_{i=1}^m g_{A,j_i}
\right]_{\leq W}.
\label{eq:app_M_finite}
\end{equation}
Here $m$ counts the number of source factors, and $\mathbb N_0^m$ denotes the set of $m$-tuples of nonnegative integers.\footnote{An element of $\mathbb N_0^m$ is an ordered tuple $(j_1,\ldots,j_m)$ with $j_i=0,1,2,\ldots$ for every $i=1,\ldots,m$.} The multi-index $\mathbf j=(j_1,\ldots,j_m)$ specifies the logarithmic power associated with each factor, with $|\mathbf j|=\sum_{i=1}^m j_i$. Since $g_{A,j}$ starts at order $j+1$ in SR, each product contributes from order $m+|\mathbf j|$. Consequently, only terms satisfying $m+|\mathbf j|\leq W$ contribute, and Eq.~\eqref{eq:app_M_finite} contains only finitely many terms at any fixed order. For example, through second order the source contributions are $\chi_{1;0}g_{A,0}$, $\chi_{1;1}g_{A,1}$, and $\chi_{2;0,0}g_{A,0}^2$.

In Sec.~\ref{sec:universal_coefficients}, we show that the coefficients $\chi_{m;\mathbf j}\equiv\chi_{m;j_1,\ldots,j_m}$ are independent of the inflationary background and are determined entirely by the free de Sitter solutions. All dependence on the inflationary background in Eq.~\eqref{eq:app_M_finite} is therefore contained in $h$ and the source coefficients $g_{A,j}$.

\subsection{The universal coefficients $\chi_{m;\mathbf j}$}
\label{sec:universal_coefficients}

We now derive the universal coefficients $\chi_{m;\mathbf j}$ entering Eq.~\eqref{eq:app_M_finite}. Our goal is to determine the super-horizon amplitude directly from the free de Sitter solutions, without calculating successive corrections to the mode functions. 

In the de Sitter limit, the Bunch--Davies mode is given by Eq.~\eqref{eq:app_BD_condition}. The corresponding free Green functions for the mode and pump equations are
\begin{equation}
G_R(x,t)=-i\operatorname{Re}[y_0(x_<)]y_0(x_>), \qquad G_P(x,t)=\frac{x_>^2-x_>^{-1}}{3x_<}.
\label{eq:app_free_kernels}
\end{equation}
Here $x_<$ and $x_>$ denote the smaller and larger arguments. The two kernels reflect the different boundary conditions of the mode and pump equations. The first combines the solution regular at $x\to0$ with the Bunch--Davies mode at large $x$, while the second combines the growing pump solution with an independent solution vanishing at $x=1$. Accordingly, $G_R$ is defined on $I_R=(0,\infty)$ and $G_P$ on $I_P=(0,1)$. Importantly, both kernels depend only on the free de Sitter solutions, so all dependence on the inflationary background enters through the source $V_A$. 

It is also useful to introduce the integral operators $K_B=G_BV_A$, with $B=R,P$, defined by
\begin{equation}
(K_B\varphi)(x)=\int_{I_B}G_B(x,t)V_A(t)\varphi(t)\,dt.
\label{eq:app_K_definition}
\end{equation}

To understand how the powers of $K_B$ are constructed, we start from its definition~\eqref{eq:app_K_definition}. Applying the same operator a second time gives
\begin{equation}
\begin{aligned}
(K_B^2\varphi)(x) &= [K_B(K_B\varphi)](x)\\
&= \int_{I_B}G_B(x,x_1)V_A(x_1)(K_B\varphi)(x_1)\,dx_1\\
&= \int_{I_B}G_B(x,x_1)V_A(x_1)\left[\int_{I_B}G_B(x_1,x_2)V_A(x_2)\varphi(x_2)\,dx_2\right]dx_1\\
&= \int_{I_B^2}G_B(x,x_1)V_A(x_1)G_B(x_1,x_2)V_A(x_2)\varphi(x_2)\,dx_1dx_2.
\end{aligned}
\end{equation}
Similarly, applying it three times gives
\begin{equation}
(K_B^3\varphi)(x)=\int_{I_B^3}G_B(x,x_1)V_A(x_1)G_B(x_1,x_2)V_A(x_2)G_B(x_2,x_3)V_A(x_3)\varphi(x_3)\,d^3x.
\end{equation}
Each application introduces one additional Green function and one source factor. The result can therefore be generalized to an arbitrary number $m$ of applications:
\begin{equation}
(K_B^m\varphi)(x)=\int_{I_B^m}\prod_{i=1}^{m}\left[G_B(x_{i-1},x_i)V_A(x_i)\right]\varphi(x_m)\,d^mx.
\end{equation}

We now take the trace of these expressions. As for an ordinary matrix, taking the trace means selecting the diagonal elements and summing over them. For an integral operator, this amounts to setting its two external arguments equal and integrating over their common value. At first order, we obtain
\begin{equation}
\operatorname{Tr}(K_B)=\int_{I_B}G_B(x_1,x_1)V_A(x_1)\,dx_1.
\end{equation}
At second order, we identify the external arguments $x$ and $x_2$ in the expression for $K_B^2$, giving
\begin{equation}
\operatorname{Tr}(K_B^2)=\int_{I_B^2}G_B(x_1,x_2)G_B(x_2,x_1)V_A(x_1)V_A(x_2)\,d^2x.
\end{equation}
Likewise, at third order,
\begin{equation}
\operatorname{Tr}(K_B^3)=\int_{I_B^3}G_B(x_1,x_2)G_B(x_2,x_3)G_B(x_3,x_1)V_A(x_1)V_A(x_2)V_A(x_3)\,d^3x.
\end{equation}
The pattern is now clear. Taking the trace connects the last Green function back to the first, producing a closed product. At arbitrary order, we therefore obtain
\begin{equation}
\operatorname{Tr}(K_B^m)=\int_{I_B^m}\prod_{i=1}^{m}\left[\frac{g_A(x_i)}{x_i^2}G_B(x_i,x_{i+1})\right]d^mx,
\label{eq:app_cyclic_trace}
\end{equation}
where $x_{m+1}=x_1$ and we have used $V_A(x)=g_A(x)/x^2$.

With these definitions and results in mind, we proceed in three steps. First, in Sec.~\ref{sec:app_explicit_calculation}, we use an identity relating the physical amplitude to the operators $K_R$ and $K_P$ to obtain an explicit, non-recursive representation of the universal coefficients. Second, we prove this identity in Sec.~\ref{sec:app_determinant_proof}. Third, we establish the regularity of the resulting expressions in Sec.~\ref{sec:app_convergence}.

\subsubsection{Explicit Calculation of the coefficients $\chi_{m;\mathbf j}$}
\label{sec:app_explicit_calculation}

We start from the following identity
\begin{equation}
J_A=\det(1-K_R)\det(1-K_P),
\label{eq:app_J_determinants}
\end{equation}
which relates the amplitude $J_A$ in Eq.~\eqref{eq:app_J} to the operators $K_R$ and $K_P$ defined in Eq.~\eqref{eq:app_K_definition}.\footnote{
Equation~\eqref{eq:app_J_determinants} is a special case of the classical relation between Jost functions and Fredholm determinants~\cite{Jost:1951zz}, previously applied to inflationary mode amplitudes in Ref.~\cite{Habib:2004hd}. For completeness, we give a self-contained proof adapted to the mode and pump boundary conditions.} This identity will be proved in Sec.~\ref{sec:app_determinant_proof} while its extension to the logarithmic sources used in the SR expansion is justified in Sec.~\ref{sec:app_convergence}. 

Starting from Eq.~\eqref{eq:app_J_determinants}, we take the logarithm of the squared modulus $\ln|J_A|^2=\ln|\det(1-K_R)|^2+\ln|\det(1-K_P)|^2$. Using the identity $\ln|z|^2=2\operatorname{Re}\ln z$, together with the standard expansion of the logarithmic determinant $\ln\det(1-K)=-\sum_{m=1}^{\infty}\operatorname{Tr}(K^m)/m$, we obtain
\begin{equation}
\ln|J_A|^2=-2\operatorname{Re}\sum_{m=1}^{\infty}\frac{1}{m}\left[\operatorname{Tr}(K_R^m)+\operatorname{Tr}(K_P^m)\right].
\label{eq:app_J_series}
\end{equation}
The traces appearing in Eq.~\eqref{eq:app_J_series} can be evaluated using Eq.~\eqref{eq:app_cyclic_trace}. Substituting their explicit expressions, we obtain
\begin{equation}
\ln|J_A|^2=-2\operatorname{Re}\sum_{m=1}^{\infty}\frac{1}{m}\left[\int_{I_R^m}\prod_{i=1}^{m}\frac{g_A(x_i)}{x_i^2}G_R(x_i,x_{i+1})\,d^mx+\int_{I_P^m}\prod_{i=1}^{m}\frac{g_A(x_i)}{x_i^2}G_P(x_i,x_{i+1})\,d^mx\right].
\label{eq:app_J_integral}
\end{equation}
We now expand the sources $g_A(x_i)$ appearing in the integrals as
\begin{equation}
g_A(x_i)=\sum_{\ell\geq0}g_{A,\ell}(\ln x_i)^\ell.
\end{equation}
To evaluate the integrals entering $\ln|J_A|^2$ in Eq.~\eqref{eq:app_J_integral}, we need to expand products of the source functions evaluated at different integration variables, such as $g_A(x_1)$, $g_A(x_1)g_A(x_2)$, and so on. For a single source insertion, we simply have $g_A(x_1)=\sum_{\ell\geq0}g_{A,\ell}(\ln x_1)^\ell$. For two source insertions we have:
\begin{equation}
g_A(x_1)g_A(x_2)=\left[\sum_{\ell\geq0}g_{A,\ell}(\ln x_1)^\ell\right]\left[\sum_{r\geq0}g_{A,r}(\ln x_2)^r\right]=\sum_{\ell,r\geq0}g_{A,\ell}g_{A,r}(\ln x_1)^\ell(\ln x_2)^r.
\end{equation}
Similarly, for three source insertions, we obtain
\begin{equation}
g_A(x_1)g_A(x_2)g_A(x_3)=\sum_{\ell,r,s\geq0}g_{A,\ell}g_{A,r}g_{A,s}(\ln x_1)^\ell(\ln x_2)^r(\ln x_3)^s.
\end{equation}
Therefore, at an arbitrary number $m$ of source insertions we have
\begin{equation}
\prod_{i=1}^{m}g_A(x_i)=\sum_{j_1,\ldots,j_m\geq0}g_{A,j_1}\cdots g_{A,j_m}(\ln x_1)^{j_1}\cdots(\ln x_m)^{j_m}.
\label{eq:app_source_product}
\end{equation}

Substituting this result into Eq.~\eqref{eq:app_J_integral} and taking the coefficients $g_{A,j_1}\cdots g_{A,j_m}$ outside the integrals (they do not depend on the integration variables), we obtain
\begin{equation}
\begin{aligned}
\ln|J_A|^2 = -2\operatorname{Re}\sum_{m=1}^{\infty}\frac{1}{m}\sum_{j_1,\ldots,j_m\geq0}g_{A,j_1}\cdots g_{A,j_m}\Bigg[&
\int_{I_R^m}\frac{(\ln x_1)^{j_1}\cdots(\ln x_m)^{j_m}}{x_1^2\cdots x_m^2}\prod_{i=1}^{m}G_R(x_i,x_{i+1})\,d^mx\\
+{}&
\int_{I_P^m}\frac{(\ln x_1)^{j_1}\cdots(\ln x_m)^{j_m}}{x_1^2\cdots x_m^2}\prod_{i=1}^{m}G_P(x_i,x_{i+1})\,d^mx
\Bigg].
\end{aligned}
\label{eq:app_comparison_1}
\end{equation}

The background dependence is now entirely contained in the coefficients $g_{A,j_1}\cdots g_{A,j_m}$, while the integrals involve only the free de Sitter Green functions and the logarithmic powers. Introducing $\mathbf j=(j_1,\ldots,j_m)$, we can therefore write
\begin{equation}
\ln|J_A|^2=\sum_{m=1}^{\infty}\sum_{j_1,\ldots,j_m\geq0}\chi_{m;\mathbf j}\,\,g_{A,j_1}\cdots g_{A,j_m}.
\label{eq:app_comparison_2}
\end{equation}
Since the source coefficients commute, only the fully symmetric part of the cyclic integrals in Eq.~\eqref{eq:app_comparison_1} contributes to the amplitude. We therefore define $\chi_{m;\mathbf j}$ by averaging over all permutations of the logarithmic indices. Here $S_m$ denotes the corresponding symmetric group, with each permutation $\tilde \pi\in S_m$ acting as $(j_1,\ldots,j_m)\mapsto(j_{\tilde\pi(1)},\ldots,j_{\tilde\pi(m)})$. This includes also permutations of coincident indices. The sum in Eq.~\eqref{eq:app_comparison_2} still runs over all ordered tuples, so no additional multiplicity factor is required. Comparing the two expressions then gives
\begin{equation}
\boxed{
\begin{aligned}\chi_{m;\mathbf j}=-\frac{2}{m\,m!}\operatorname{Re}\sum_{\pi\in S_m}\Bigg[
&\int_{I_R^m}\frac{(\ln x_1)^{j_{\pi(1)}}\cdots(\ln x_m)^{j_{\pi(m)}}}{x_1^2\cdots x_m^2}\prod_{i=1}^{m}G_R(x_i,x_{i+1})\,d^mx\\
&+\int_{I_P^m}\frac{(\ln x_1)^{j_{\pi(1)}}\cdots(\ln x_m)^{j_{\pi(m)}}}{x_1^2\cdots x_m^2}\prod_{i=1}^{m}G_P(x_i,x_{i+1})\,d^mx\Bigg].
\end{aligned}
}
\label{eq:app_chi_direct}
\end{equation}
Equation~\eqref{eq:app_chi_direct} provides the desired coefficients at arbitrary order. The multidimensional nature of these integrals is not an artifact of the calculation. At order $m$ in the source, the perturbations receive $m$ contributions from the inflationary background, connected by the free Green functions. Each contribution can occur at a different value of $x$, so we must integrate over all possible values of these $m$ variables. This naturally leads to an $m$-dimensional integral. Importantly, however, the integrals contain only the free Green functions defined in Eq.~\eqref{eq:app_free_kernels}. No perturbatively corrected mode functions or lower-order amplitudes are required. This explicit, non-recursive integral representation is the main physical result of this calculation.

Although this representation is physically transparent, directly evaluating multidimensional integrals becomes increasingly expensive at higher orders. We therefore show that the integrals~\eqref{eq:app_chi_direct} can be evaluated in terms of Gamma functions and explicit series, allowing the universal coefficients to be obtained through successive derivatives with respect to auxiliary exponents. This alternative representation is particularly useful for computational implementation, as it replaces multidimensional integrations with the evaluation of known functions and their derivatives.

Since the Green functions depend on the smaller and larger of their arguments, we divide the integration domains into sectors corresponding to all possible orderings of $x_1,\ldots,x_m$. We first consider the sector $0<x_1<\cdots<x_m$, while the remaining sectors follow by permuting the integration variables. We also temporarily replace the logarithmic factors $(\ln x_i)^{j_i}$ by auxiliary powers $x_i^{\mu_i}$. The original logarithms can be recovered at the end using the identity $\left.\partial_\mu^j x^\mu\right|_{\mu=0}=(\ln x)^j$. With this replacement, the expression inside the real part of Eq.~\eqref{eq:app_chi_direct} becomes
\begin{equation}
\begin{aligned}\mathcal C_m(\boldsymbol\mu)=-\frac{2}{m\,m!}\sum_{\pi\in S_m}\Bigg[
&\int_{I_R^m}\frac{x_1^{\mu_{\pi(1)}}\cdots x_m^{\mu_{\pi(m)}}}{x_1^2\cdots x_m^2}\prod_{i=1}^{m}G_R(x_i,x_{i+1})\,d^mx\\
&+\int_{I_P^m}\frac{x_1^{\mu_{\pi(1)}}\cdots x_m^{\mu_{\pi(m)}}}{x_1^2\cdots x_m^2}\prod_{i=1}^{m}G_P(x_i,x_{i+1})\,d^mx\Bigg].
\end{aligned}
\label{eq:app_C_auxiliary}
\end{equation}

Once the ordering of the variables is fixed, we know which argument is smaller and which is larger in each Green function. We can therefore substitute their explicit expressions from Eq.~\eqref{eq:app_free_kernels} and expand the resulting products. For the pump, this produces a finite sum of ordered integrals containing only powers $x_i^{a_i-1}$, while for the mode the corresponding terms also contain exponential factors $e^{-\beta_i x_i}$. Consequently, $\mathcal C_m(\boldsymbol\mu)$ reduces to a finite sum of ordered integrals of these two types
\begin{equation}
\underbrace{\int_{0<x_1<\cdots<x_m<\infty}\prod_{i=1}^{m}x_i^{a_i-1}e^{-\beta_i x_i}\,d^mx}_{\text{mode integral}}
+
\underbrace{\int_{0<x_1<\cdots<x_m<1}\prod_{i=1}^{m}x_i^{a_i-1}\,d^mx}_{\text{pump integral}},
\label{eq:pump_and_mode_integrals}
\end{equation}
summed over all possible orderings of the variables. 
Here the exponents $a_i$ collect the powers arising from the Green functions and the auxiliary factors $x_i^{\mu_i-2}$, while $\beta_i$ specifies the exponential factors. We can now evaluate these two integrals separately. 

We start with the simpler pump integral in Eq.~\eqref{eq:pump_and_mode_integrals}. The ordering allows us to perform the integrations successively, starting from the smallest variable. For example, with two variables, we have
\begin{equation}
\int_{0<x_1<x_2<t}\prod_{i=1}^{2}x_i^{a_i-1}\,d^2x=\int_0^t x_2^{a_2-1}\left[\int_0^{x_2}x_1^{a_1-1}\,dx_1\right]dx_2=\frac{1}{a_1}\int_0^t x_2^{a_1+a_2-1}\,dx_2=\frac{t^{a_1+a_2}}{a_1(a_1+a_2)}.
\end{equation}
The first integration produces the denominator $a_1$, while the second produces $a_1+a_2$. At higher orders, each successive integration adds the corresponding exponent to those already accumulated. Defining $A_j=\sum_{i=1}^{j}a_i$, we obtain
\begin{equation}
\boxed{
\text{pump integral}=\frac{t^{A_m}}{\prod_{j=1}^{m}A_j}\overset{t=1}{=}\prod_{j=1}^{m}\frac{1}{A_j}.
}
\label{eq:app_ordered_pump}
\end{equation}
The last equality applies to the pump contribution, for which all integration variables lie in $(0,1)$, so the upper integration limit is $t=1$. The result is initially valid for $\operatorname{Re}A_j>0$.

We now turn to the mode contribution in Eq.~\eqref{eq:pump_and_mode_integrals}. Unlike the pump, the free mode contains exponential factors inherited from $y_0=(1+i/x)e^{ix}$. 
The idea is to reduce the internal integrations to the same elementary power integrals used for the pump. To achieve this, we expand the exponentials associated with the first $m-1$ variables, while keeping the exponential of the largest variable, $t=x_m$, unchanged. The internal integrations can then be performed directly, while the remaining integral over $t$ is evaluated using the Gamma function. For example, when $m=2$, we expand $e^{-\beta_1x_1}=\sum_{k\geq0}(-\beta_1)^kx_1^k/k!$ and integrate over $x_1$, obtaining
\begin{equation}
\begin{aligned}
\int_{0<x_1<x_2<\infty}\prod_{i=1}^{2}x_i^{a_i-1}e^{-\beta_i x_i}\,d^2x
&=\sum_{k=0}^{\infty}\frac{(-\beta_1)^k}{k!}\int_0^\infty t^{a_2-1}e^{-\beta_2t}\left[\int_0^t x_1^{a_1+k-1}\,dx_1\right]dt\\
&=\sum_{k=0}^{\infty}\frac{(-\beta_1)^k}{k!(a_1+k)}\int_0^\infty t^{a_1+a_2+k-1}e^{-\beta_2t}\,dt\\
&=\sum_{k=0}^{\infty}\frac{(-\beta_1)^k}{k!(a_1+k)}\frac{\Gamma(a_1+a_2+k)}{\beta_2^{a_1+a_2+k}}.
\end{aligned}
\end{equation}
where in the last step we used the Gamma-function identity $\int_0^\infty t^{z-1}e^{-\beta t}\,dt=\Gamma(z)\beta^{-z}$. Thus, the mode contribution can also be evaluated using successive integrations, without introducing any additional change of variables. The same procedure extends to arbitrary $m$. We first separate the integration over the largest variable, $t=x_m$, from the remaining integrations:
\begin{equation}
\text{mode integral}=\int_0^\infty t^{a_m-1}e^{-\beta_m t}\left[\int_{0<x_1<\cdots<x_{m-1}<t}\prod_{i=1}^{m-1}x_i^{a_i-1}e^{-\beta_i x_i}\,d^{m-1}x\right]dt.
\label{eq:mode_int_expansion}
\end{equation}
Expanding the exponentials gives
\begin{equation}
\prod_{i=1}^{m-1}e^{-\beta_i x_i}=\sum_{k_1,\ldots,k_{m-1}\geq0}\prod_{i=1}^{m-1}\frac{(-\beta_i)^{k_i}}{k_i!}x_i^{k_i}.
\end{equation}
Substituting this expansion into the internal integral in the square brackets of Eq.~\eqref{eq:mode_int_expansion}, we obtain
\begin{equation}
\begin{aligned}
\int_{0<x_1<\cdots<x_{m-1}<t}\prod_{i=1}^{m-1}x_i^{a_i-1}e^{-\beta_i x_i}\,d^{m-1}x
&=\sum_{k_1,\ldots,k_{m-1}\geq0}\left(\prod_{i=1}^{m-1}\frac{(-\beta_i)^{k_i}}{k_i!}\right)\int_{0<x_1<\cdots<x_{m-1}<t}\prod_{i=1}^{m-1}x_i^{a_i+k_i-1}\,d^{m-1}x\\
&=\sum_{k_1,\ldots,k_{m-1}\geq0}\left(\prod_{i=1}^{m-1}\frac{(-\beta_i)^{k_i}}{k_i!}\right)\frac{t^{A_{m-1}+K_{m-1}}}{\prod_{j=1}^{m-1}(A_j+K_j)},
\end{aligned}
\end{equation}
where $K_j=\sum_{i=1}^{j}k_i$, and the last equality follows from Eq.~\eqref{eq:app_ordered_pump}. The indices $k_i$ belong to the exponential expansion, so the SR truncation at order $W$ does not truncate these sums. We can now substitute this result back into the original integral over $t$. Writing $A=A_m$, we find
\begin{equation}
\text{mode integral}=\sum_{k_1,\ldots,k_{m-1}\geq0}\frac{\prod_{i=1}^{m-1}(-\beta_i)^{k_i}/k_i!}{\prod_{j=1}^{m-1}(A_j+K_j)}\int_0^\infty t^{A+K_{m-1}-1}e^{-\beta_m t}\,dt.
\label{eq:app_ordered_mode_reduced}
\end{equation}
The remaining integral is evaluated using the Gamma-function identity $\int_0^\infty t^{z-1}e^{-\beta t}\,dt=\Gamma(z)\beta^{-z}$, finally giving
\begin{equation}
\boxed{
\text{mode integral}=\sum_{k_1,\ldots,k_{m-1}\geq0}\frac{\Gamma(A+K_{m-1})}{\beta_m^{A+K_{m-1}}}\frac{\prod_{i=1}^{m-1}(-\beta_i)^{k_i}/k_i!}{\prod_{j=1}^{m-1}(A_j+K_j)}.
}
\label{eq:app_ordered_mode}
\end{equation}
For $m=1$, the sums and products over internal variables are absent, and this expression reduces directly to the Gamma-function integral.
The series in Eq.~\eqref{eq:app_ordered_mode} is initially evaluated for $\operatorname{Re}A_j>0$ and $\operatorname{Re}\beta_m>\sum_{i<m}|\beta_i|$, which justify interchanging the sums and integrals. For the physical mode terms, we then continue to $\beta_i=\delta-2i\sigma_i$, with $\sigma_i\in\{-1,0,1\}$ and $\delta>0$. Nonoscillating terms therefore have $\beta_i=\delta$, rather than zero. The complex powers are continued from $\operatorname{Re}\beta_i>0$, fixing the outgoing prescription. All kernel terms and ordered sectors, together with the pump contribution, are combined before taking $\delta\to0^+$ at generic auxiliary exponents. The resulting expression is then continued to $\boldsymbol\mu=0$ as established in Sec.~\ref{sec:app_convergence}.

We have therefore obtained explicit expressions for both the pump and mode integrals for arbitrary $m$. The convergence of these expressions and their continuation to the physical sources are discussed in Sec.~\ref{sec:app_convergence}. We can now recover the logarithmic powers that were replaced by the auxiliary exponents $\mu_i$. Recall that each factor $x_i^{\mu_i}$ was introduced to generate the corresponding power $(\ln x_i)^{j_i}$ through differentiation. For example, when $m=2$, differentiating once with respect to $\mu_1$ and twice with respect to $\mu_2$ gives $\left.\partial_{\mu_1}\partial_{\mu_2}^2(x_1^{\mu_1}x_2^{\mu_2})\right|_{\boldsymbol\mu=0}=(\ln x_1)(\ln x_2)^2$. The same procedure applies to arbitrary $m$ and any combination of logarithmic powers. Since the Green functions do not depend on the auxiliary exponents, these derivatives reproduce exactly the logarithmic factors appearing in Eq.~\eqref{eq:app_chi_direct}. We can therefore first evaluate $\mathcal C_m(\boldsymbol\mu)$ using the explicit expressions for the ordered integrals derived above, and then differentiate the result with respect to the auxiliary exponents. The universal coefficients are thus obtained as
\begin{equation}
\boxed{
\chi_{m;\mathbf j}=\operatorname{Re}\left[\left.\frac{\partial^{j_1}}{\partial\mu_1^{j_1}}\cdots\frac{\partial^{j_m}}{\partial\mu_m^{j_m}}\mathcal C_m(\boldsymbol\mu)\right|_{\boldsymbol\mu=0}\right].
}
\label{eq:app_chi_log_derivatives}
\end{equation}
Here $\mathcal C_m(\boldsymbol\mu)$ is evaluated by summing the explicit ordered-integral expressions over all possible orderings of the variables and averaging over permutations of the auxiliary exponents as in Eq.~\eqref{eq:app_C_auxiliary}. This makes the coefficients extracted in Eq.~\eqref{eq:app_chi_log_derivatives} fully symmetric in $\mathbf j$. The mode and pump contributions are combined before taking the limit $\boldsymbol\mu\to0$, using the common continuation established in Sec.~\ref{sec:app_convergence}.

Equations~\eqref{eq:app_ordered_pump}, \eqref{eq:app_ordered_mode}, and~\eqref{eq:app_chi_log_derivatives} determine the universal coefficients directly from the free de Sitter solutions, without requiring perturbatively corrected mode functions or lower-order amplitudes. An equivalent representation without auxiliary derivatives can be obtained using Bell polynomials and finite combinatorial sums. However, we stick to the derivative form for compactness.

\subsubsection{Proof of the determinant identity in Eq.~\eqref{eq:app_J_determinants}}
\label{sec:app_determinant_proof}

We now establish Eq.~\eqref{eq:app_J_determinants}, which relates the physical super-horizon amplitude to the determinants of the integral operators constructed from the free mode and pump Green functions. We first consider a smooth source $V_A$ that vanishes outside the interval $a<x<b$, with $0<a<1<b$. This ensures that the source contributes only within this finite interval and that the solutions follow the free equations outside it. The extension to the logarithmic sources is addressed in Sec.~\ref{sec:app_convergence}.

We first note that the free Green functions defined in Eq.~\eqref{eq:app_free_kernels} have the common form
\begin{equation}
G_B(x,t)=\frac{u_B(x_<)v_B(x_>)}{w_B},
\end{equation}
where $w_B\equiv W[u_B,v_B]=u_Bv_B'-u_B'v_B$ is the Wronskian. For the mode, $u_R=\operatorname{Re}y_0$, $v_R=y_0$, and $w_R=i$. For the pump, $u_P=1/x$, $v_P=(x^2-x^{-1})/3$, and $w_P=1$.

A second key observation is that the Green functions can be rewritten in two equivalent ways. For the pump, we use
\begin{equation}
G_P(x,t)=\frac{u_P(x)v_P(t)}{w_P}+\Theta(x-t)\frac{u_P(t)v_P(x)-u_P(x)v_P(t)}{w_P},
\end{equation}
while for the mode we use
\begin{equation}
G_R(x,t)=\frac{v_R(x)u_R(t)}{w_R}+\Theta(t-x)\frac{u_R(x)v_R(t)-v_R(x)u_R(t)}{w_R}.
\end{equation}
Here $\Theta$ denotes the Heaviside step function. Both expressions are exact identities obtained by considering separately the cases $x<t$ and $x>t$. The difference is that the second term involves only $t<x$ for the pump and only $t>x$ for the mode. This allows us to construct the two solutions starting from the boundaries where their physical behavior is known.

Although the two decompositions appear different, they have the same mathematical structure. To treat them together, we denote by $s_B$ the free solution whose boundary behavior we want to preserve, and by $r_B$ the other independent free solution, with the index $B=(P,R)$. For the pump, we take $s_P=u_P$ and $r_P=v_P$, while for the mode we take $s_R=v_R$ and $r_R=u_R$. We also introduce a sign $\eta_B$, with $\eta_P=+1$ and $\eta_R=-1$, to distinguish the two integration directions. Both Green functions can then be written as
\begin{equation}
G_B(x,t)=\frac{s_B(x)r_B(t)}{w_B}+\Theta\!\left((x-t)\eta_B\right)\frac{s_B(t)r_B(x)-s_B(x)r_B(t)}{w_B}.
\label{eq:app_kernel_decomposition}
\end{equation}
For $\eta_B=+1$, the second term contributes only when $t<x$, whereas for $\eta_B=-1$ it contributes only when $t>x$. Equation~\eqref{eq:app_kernel_decomposition} therefore reproduces the two decompositions above.

We now substitute Eq.~\eqref{eq:app_kernel_decomposition} into the definition of the integral operator $K_B$ in Eq.~\eqref{eq:app_K_definition}. For the source considered here, the integration domains are $I_R=(a,b)$ and $I_P=(a,1)$. We obtain
\begin{equation}
\begin{aligned}
(K_B\varphi)(x)&=\int_{I_B}G_B(x,t)V_A(t)\varphi(t)\,dt\\
&=\int_{I_B}\left[\frac{s_B(x)r_B(t)}{w_B}+\Theta\!\left((x-t)\eta_B\right)\frac{s_B(t)r_B(x)-s_B(x)r_B(t)}{w_B}\right]V_A(t)\varphi(t)\,dt\\
&=\frac{s_B(x)}{w_B}\int_{I_B}r_B(t)V_A(t)\varphi(t)\,dt+\int_{I_B}\Theta\!\left((x-t)\eta_B\right)\frac{s_B(t)r_B(x)-s_B(x)r_B(t)}{w_B}V_A(t)\varphi(t)\,dt.
\end{aligned}
\end{equation}
In the last step, we have taken $s_B(x)$ outside the first integral, since it does not depend on $t$. We can now identify two separate contributions. The first is proportional to $s_B(x)$, with a coefficient that depends on $\varphi$ but not on $x$. We denote this coefficient by
\begin{equation}
\ell_B[\varphi]=\frac{1}{w_B}\int_{I_B}r_B(t)V_A(t)\varphi(t)\,dt.
\label{eq:ell_B_def}
\end{equation}
The second contribution defines the operator
\begin{equation}
(T_B\varphi)(x)=\int_{I_B}\Theta\!\left(\eta_B(x-t)\right)\frac{s_B(t)r_B(x)-s_B(x)r_B(t)}{w_B}V_A(t)\varphi(t)\,dt.
\label{eq:app_triangular_operator}
\end{equation}
For the pump, this operator integrates over $a<t<x$, while for the mode it integrates over $x<t<b$. The original operator can therefore be written as
\begin{equation}
K_B=T_B+s_B\ell_B.
\label{eq:app_operator_decomposition}
\end{equation}

We now introduce a function $f_B$ through the integral equation
\begin{equation}
f_B=s_B+T_Bf_B.
\label{eq:app_f_integral}
\end{equation}
Our first goal is to show that $f_B$ satisfies the corresponding perturbed differential equation in Eq.~\eqref{eq:app_mode_pump_equations}. Substituting the explicit expression of $T_B$, we have
\begin{equation}
f_B(x)=s_B(x)+\int_{I_B}\Theta\!\left(\eta_B(x-t)\right)\frac{s_B(t)r_B(x)-s_B(x)r_B(t)}{w_B}V_A(t)f_B(t)\,dt.
\end{equation}
Differentiating once with respect to $x$, we obtain
\begin{equation}
f_B'(x)=s_B'(x)+\int_{I_B}\Theta\!\left(\eta_B(x-t)\right)\frac{s_B(t)r_B'(x)-s_B'(x)r_B(t)}{w_B}V_A(t)f_B(t)\,dt.
\end{equation}
No additional term appears from the integration boundary because the numerator of the original kernel vanishes when $t=x$. Differentiating a second time gives
\begin{equation}
f_B''(x)=s_B''(x)+V_A(x)f_B(x)+\int_{I_B}\Theta\!\left(\eta_B(x-t)\right)\frac{s_B(t)r_B''(x)-s_B''(x)r_B(t)}{w_B}V_A(t)f_B(t)\,dt.
\end{equation}
This time, differentiation of the integration boundary produces the term $V_A(x)f_B(x)$. Indeed, the Wronskian of the two free solutions satisfies $W[s_B,r_B]=\eta_Bw_B$, so the sign from differentiating the Heaviside function combines with the Wronskian to give $\eta_B^2=1$. The result is therefore the same for the mode and pump.

Moreover, $s_B$ and $r_B$ satisfy the corresponding free differential equation. Combining this property with Eq.~\eqref{eq:app_f_integral}, we recover
\begin{equation}
f_R''+\left(1-\frac{2}{x^2}\right)f_R=V_Af_R,\qquad f_P''-\frac{2}{x^2}f_P=V_Af_P.
\end{equation}
These are precisely the perturbed mode and pump equations introduced in Eq.~\eqref{eq:app_mode_pump_equations}. The physical boundary conditions also follow directly from the integral equation. For the pump, $\eta_P=+1$, so the integral vanishes below the source and $f_P=s_P=u_P=1/x$ for $x<a$. For the mode, $\eta_R=-1$, so the integral vanishes above the source and $f_R=s_R=v_R=y_0$ for $x>b$. Therefore, $f_R$ is exactly the Bunch--Davies solution $y_A$, while $f_P$ is the growing pump solution before imposing its normalization at $x=1$.

We now use Eq.~\eqref{eq:app_f_integral} to simplify $\det(1-K_B)$. From $f_B=s_B+T_Bf_B$, we immediately obtain $(1-T_B)f_B=s_B$. Recall also that Eq.~\eqref{eq:app_operator_decomposition} gives $K_B=T_B+s_B\ell_B$. Combining these two relations, we can write
\begin{equation}
1-K_B=1-T_B-s_B\ell_B=1-T_B-(1-T_B)f_B\ell_B=(1-T_B)(1-f_B\ell_B).
\label{eq:app_determinant_factorization}
\end{equation}
We have therefore expressed $1-K_B$ as a product of two simpler operators. Since the determinant of a product is the product of the determinants, we have
\begin{equation}
\det(1-K_B)=\det(1-T_B)\det(1-f_B\ell_B).
\label{eq:det_combined}
\end{equation}

We first consider $\det(1-T_B)$. From Eq.~\eqref{eq:app_triangular_operator}, $T_B$ integrates only in one direction: over $t<x$ for the pump and over $t>x$ for the mode. This property has an important consequence when computing the traces of its powers. For example, $\operatorname{Tr}(T_B^2)$ would require two integration variables satisfying both $x_1>x_2$ and $x_2>x_1$, which is impossible. Similarly, $\operatorname{Tr}(T_B^m)$ requires a sequence of variables to be strictly increasing or decreasing and simultaneously return to its starting point. The trace also vanishes for $m=1$, since the kernel is zero when its two arguments coincide. Thus, $\operatorname{Tr}(T_B^m)=0$ for every $m\geq1$. Using the logarithmic determinant expansion introduced above, we immediately find
\begin{equation}
\det(1-T_B)=1.
\label{Eq:det_1}
\end{equation}

We now consider the second determinant $\det(1-f_B\ell_B)$. The operator $f_B\ell_B$ acts on an arbitrary function $\varphi$ as $(f_B\ell_B)\varphi=f_B\ell_B[\varphi]$. This means that, regardless of the function $\varphi$, the result is always proportional to $f_B$. In particular, applying this operator to $f_B$ itself gives $(f_B\ell_B)f_B=\ell_B[f_B]f_B$. Therefore, $f_B$ is an eigenfunction with eigenvalue $\ell_B[f_B]$. Since the operator maps every function onto a multiple of $f_B$, it has no other nonzero eigenvalues. Its determinant is consequently
\begin{equation}
\det(1-f_B\ell_B)=1-\ell_B[f_B].
\label{Eq:det_2}
\end{equation}

Combining Eqs.~\eqref{Eq:det_1} and~\eqref{Eq:det_2} in Eq.~\eqref{eq:det_combined}, we obtain
\begin{equation}
\det(1-K_B)=1-\ell_B[f_B].
\label{eq:app_determinant_integral}
\end{equation}

We can now relate the two determinants directly to the physical mode and pump solutions. We begin with the pump. Evaluating Eq.~\eqref{eq:app_f_integral} at $x=1$, and using $s_P(1)=u_P(1)=1$, $r_P(1)=v_P(1)=0$, and $w_P=1$, we obtain
\begin{equation}
f_P(1)=1-\frac{1}{w_P}\int_a^1r_P(t)V_A(t)f_P(t)\,dt=1-\ell_P[f_P].
\end{equation}
Comparing this expression with Eq.~\eqref{eq:app_determinant_integral}, we immediately find $\det(1-K_P)=f_P(1)$. Since $f_P=1/x$ below the source and the physical pump is normalized by $Z_A=f_P/f_P(1)$, we have $Z_A=1/[xf_P(1)]$ for $x<a$. Consequently,
\begin{equation}
\det(1-K_P)=f_P(1)=\frac{1}{xZ_A(x)},\qquad x<a.
\label{eq:app_pump_determinant}
\end{equation}

For the mode, we recall that $f_R=y_A$, the Bunch--Davies solution. Below the source, where $x<a$, the Heaviside function in Eq.~\eqref{eq:app_f_integral} equals unity throughout the integration domain. We therefore obtain
\begin{equation}
\begin{aligned}
y_A(x)&=s_R(x)+\int_a^b\frac{s_R(t)r_R(x)-s_R(x)r_R(t)}{w_R}V_A(t)y_A(t)\,dt\\
&=\left[1-\ell_R[y_A]\right]s_R(x)+\frac{r_R(x)}{w_R}\int_a^bs_R(t)V_A(t)y_A(t)\,dt.
\end{aligned}
\end{equation}
Using Eq.~\eqref{eq:app_determinant_integral}, the coefficient multiplying $s_R(x)=y_0(x)$ is precisely $\det(1-K_R)$. We can therefore write
\begin{equation}
y_A(x)=\det(1-K_R)y_0(x)+\frac{\operatorname{Re}y_0(x)}{w_R}\int_a^by_0(t)V_A(t)y_A(t)\,dt,\qquad x<a.
\end{equation}

In the super-horizon limit, $xy_0(x)\to i$, while $x\operatorname{Re}y_0(x)\to0$. Multiplying the previous equation by $x$ and taking $x\to0$, the second contribution therefore vanishes, leaving $\lim_{x\to0}xy_A(x)=i\det(1-K_R)$. Consequently,
\begin{equation}
\det(1-K_R)=-i\lim_{x\to0}xy_A(x).
\label{eq:app_mode_determinant}
\end{equation}

Finally, combining Eqs.~\eqref{eq:app_pump_determinant} and~\eqref{eq:app_mode_determinant} with the definition of the physical amplitude~\eqref{eq:app_J}, we find
\begin{equation}
J_A=-i\lim_{x\to0}\frac{y_A}{Z_A}=\left[-i\lim_{x\to0}xy_A(x)\right]f_P(1)=\det(1-K_R)\det(1-K_P).
\end{equation}
This establishes Eq.~\eqref{eq:app_J_determinants} for the regulated source.

\subsubsection{Convergence and regularity}
\label{sec:app_convergence}

In the previous subsection, we proved the determinant identity in Eq.~\eqref{eq:app_J_determinants} assuming that the source vanishes near $x=0$. However, the logarithmic sources in Eq.~\eqref{eq:app_M_finite} do not satisfy this assumption. We must therefore establish whether the determinant representation remains valid when these sources extend to the origin. The main difficulty is that the individual mode and pump integrals in Eq.~\eqref{eq:app_chi_direct} can diverge, even though their combination describes a physical amplitude that may remain finite.

To address this problem, we introduce auxiliary sources
\begin{equation}
g_A(x)=\sum_a\lambda_a x^{\mu_a}.
\label{eq:app_auxiliary_source}
\end{equation}
These sources allow us to recover the required logarithmic powers by differentiation, since $\partial_{\mu_a}^j x^{\mu_a}|_{\mu_a=0}=(\ln x)^j$. Our strategy is therefore to establish the determinant identity for values of $\mu_a$ where the integrals converge, and then determine whether the result can be extended to $\boldsymbol\mu=0$. We initially keep the upper integration limit finite, so that we can focus on the behavior near $x=0$.

We begin by determining when the determinant integrals converge at the lower integration limit. Recall that the operators in Eq.~\eqref{eq:app_K_definition} contain the product $G_B(x,t)V_A(t)$, with $V_A(t)=g_A(t)/t^2$ according to Eq.~\eqref{eq:app_mode_pump_equations}. The first-order contribution to the determinant involves $\operatorname{Tr}K_B$, obtained by evaluating the Green function at $t=x$ and integrating over $x$. We can therefore identify the strongest possible divergence by examining the diagonal Green functions.

Consider first the pump. From Eq.~\eqref{eq:app_free_kernels}, using $u_P=1/x$, $v_P=(x^2-x^{-1})/3$, and $w_P=1$, we obtain
\begin{equation}
G_P(x,x)=\frac{u_P(x)v_P(x)}{w_P}=\frac{x-x^{-2}}{3}.
\end{equation}
For a single auxiliary source $g_A=x^\mu$, we have $V_A=x^{\mu-2}$, so the corresponding first-order integrand becomes
\begin{equation}
G_P(x,x)V_A(x)=\frac{1}{3}\left(x^{\mu-1}-x^{\mu-4}\right).
\end{equation}
The second term dominates as $x\to0$, and its integral converges only for $\operatorname{Re}\mu>3$. The mode imposes a weaker condition: using $u_R=\operatorname{Re}y_0=-x^2/3+\mathcal O(x^4)$ and $v_R=y_0=i/x+\mathcal O(x)$, we find $G_R(x,x)\sim-x/3$. Its first-order integrand therefore behaves as $x^{\mu-1}$, which converges already for $\operatorname{Re}\mu>0$.

We must also verify that the condition obtained from the pump is sufficient beyond first order. Recall that Eq.~\eqref{eq:app_chi_direct} contains products of Green functions whose arguments form a closed sequence, with $x_{m+1}=x_1$. On a finite interval, both free Green functions are bounded by a constant times $1/(xt)$. Consequently, each integration variable contributes at most two inverse powers from the Green functions and two more from the potential. The resulting integrands are bounded by products of $x_i^{\operatorname{Re}\mu_i-4}$, which are integrable when all auxiliary exponents satisfy $\operatorname{Re}\mu_a>3$.

We have therefore identified a range of auxiliary exponents in which the lower cutoff can be removed at every finite perturbative order. Since the determinant identity was already established with a finite lower cutoff, it continues to reproduce the physical amplitude in this convergence region.

This resolves the first difficulty, but not the original problem. The logarithmic sources correspond to $\boldsymbol\mu=0$, which lies outside the convergence region $\operatorname{Re}\mu_a>3$ we have just identified. We must therefore determine whether the physical amplitude can be continued from $\operatorname{Re}\mu_a>3$ to a neighborhood of $\boldsymbol\mu=0$, even when the separate determinant integrals cease to converge.

We first examine the pump. From Eq.~\eqref{eq:app_mode_pump_equations}, its differential equation is $Z_A''=(2+g_A)Z_A/x^2$. Introducing $D=d/d\ln x$ and defining $F_A=xZ_A$, we can rewrite the pump equation as
\begin{equation}
D(D-3)F_A=g_AF_A.
\label{eq:app_pump_linear}
\end{equation}
This equation is linear in $F_A$ and describes the same pump solution, with the normalization $F_A(1)=1$. In the free limit, $Z_A=1/x$, so $F_A=1$. We now solve Eq.~\eqref{eq:app_pump_linear} perturbatively in the amplitudes $\lambda_a$ defined in Eq.~\eqref{eq:app_auxiliary_source}. Writing $F_A=1+F_A^{(1)}+F_A^{(2)}+\cdots$, the first-order equation is $D(D-3)F_A^{(1)}=g_A$. Since $D(D-3)x^{\mu_a}=\mu_a(\mu_a-3)x^{\mu_a}$, imposing $F_A(1)=1$ and excluding the independent homogeneous solution proportional to $x^3$ gives
\begin{equation}
F_A^{(1)}(x)=\sum_a\lambda_a\frac{x^{\mu_a}-1}{\mu_a(\mu_a-3)}.
\label{eq:app_normalized_pump_continuation}
\end{equation}
The homogeneous contribution proportional to $x^3$ is absent because we follow the growing pump solution selected by removing the lower cutoff in the initial convergence region. Although Eq.~\eqref{eq:app_normalized_pump_continuation} contains $\mu_a$ in the denominator, the numerator also vanishes when $\mu_a\to0$. In fact, $(x^{\mu_a}-1)/[\mu_a(\mu_a-3)]$ approaches $-\ln x/3$, so the first-order correction remains regular.

To establish regularity beyond first order, we introduce $q=D\ln F_A$. Dividing Eq.~\eqref{eq:app_pump_linear} by $F_A$ gives
\begin{equation}
(D-3)q+q^2=g_A.
\label{eq:app_pump_log_derivative}
\end{equation}
At each source order, the terms driving $q$ are sums of powers $x^S$, where $S$ is a sum of auxiliary exponents. Solving for each power introduces only a denominator $S-3$, which remains nonzero near $\boldsymbol\mu=0$ at any fixed order. The independent homogeneous term proportional to $x^3$ is absent on the branch selected in the initial convergence region. The normalization $F_A(1)=1$ then gives $\ln F_A(x)=\int_1^x q(t)\,dt/t$. This introduces factors $(x^S-1)/S$, regular at $S=0$, and establishes the regularity of $\ln F_A$ and hence of the normalized pump at every fixed $x>0$.

We have now established that the pump admits a regular continuation to the auxiliary exponents required for the logarithmic sources. Nevertheless, our final quantity is not the pump itself, but the physical amplitude $J_A=-i\lim_{x\to0}y_A/Z_A$. The regularity of $Z_A$ at fixed $x$ does not automatically guarantee that this limit exists. In particular, both $y_A$ and $Z_A$ can diverge separately as $x\to0$. We must therefore show that their ratio remains finite and that its perturbative coefficients are regular near $\boldsymbol\mu=0$.

To address this remaining problem, we introduce $Y_A=y_A/Z_A$ and derive an equation directly for this ratio. Substituting $y_A=Z_AY_A$ into the mode equation and using the pump equation in Eq.~\eqref{eq:app_mode_pump_equations}, the explicit dependence on $g_A$ cancels, giving
\begin{equation}
(Z_A^2Y_A')'=-Z_A^2Y_A.
\end{equation}
We integrate this equation twice, using the values of $Y_A$ and its first derivative at $x=1$, and obtain
\begin{equation}
Y_A(x)=Y_A(1)-Y_A'(1)\int_x^1\frac{ds}{Z_A^2(s)}-\int_x^1\mathcal K(x,t)Y_A(t)\,dt,
\label{eq:app_ratio_integral}
\end{equation}
where
\begin{equation}
\mathcal K(x,t)=Z_A^2(t)\int_x^t\frac{ds}{Z_A^2(s)}.
\label{eq:app_ratio_kernel}
\end{equation}
Equation~\eqref{eq:app_ratio_integral} is equivalent to the differential equation for $Y_A$, but it makes the problem easier to study because now we can determine whether $Y_A$ has a finite super-horizon limit by examining the integrals on its right-hand side.

We first consider the free solution $Z_A=1/x$. Substituting it into Eq.~\eqref{eq:app_ratio_kernel}, we obtain
\begin{equation}
\mathcal K_0(x,t)=\frac{t^3-x^3}{3t^2}.
\end{equation}
For $0<x<t$, this kernel satisfies $0\leq\mathcal K_0(x,t)\leq t/3$. Its integral is therefore finite as $x\to0$. The same is true of the first integral in Eq.~\eqref{eq:app_ratio_integral}, since $Z_A^{-2}=x^2$ in the free limit. Thus, the integral representation is well defined at the lower integration limit when the source vanishes.

We must now verify that the perturbative corrections do not spoil this property. From the perturbative solution of Eq.~\eqref{eq:app_pump_linear}, we have already established that the corrections to the pump involve auxiliary powers $x^S$ and, when the exponents approach zero, powers of $\ln x$. At every finite perturbative order, these corrections generate corresponding powers and logarithms in $Z_A^{-2}$ and in the kernel of Eq.~\eqref{eq:app_ratio_kernel}.

The important point is that the free kernel is proportional to a positive power of $t$ near the origin. For any fixed perturbative order, we can choose the auxiliary exponents sufficiently close to zero so that the additional powers do not remove this suppression. More precisely, the kernel coefficients can be bounded by $Ct^{1-\eta}(1+|\ln t|)^r$, with $0<\eta<1$ and finite $r$. This function is integrable at $t=0$, and the same property holds for the coefficients of $Z_A^{-2}$. Writing $b(t)=Ct^{1-\eta}(1+|\ln t|)^r$ for a common bound, the ordered integrals generated by $\ell$ successive substitutions in Eq.~\eqref{eq:app_ratio_integral} are bounded by $B^\ell/\ell!$, where $B=\int_0^1 b(t)\,dt<\infty$. This ensures uniform convergence at every fixed source order and a finite limit as $x\to0$.

Furthermore, differentiating with respect to the auxiliary exponents introduces only additional powers of $\ln t$, which do not spoil the integrability of these expressions. The values of $Y_A(1)$ and $Y_A'(1)$ are also regular in the auxiliary exponents for a finite upper cutoff. We have therefore shown that the physical super-horizon amplitude is regular near $\boldsymbol\mu=0$ at every finite perturbative order, including after the derivatives required to generate the logarithmic sources are taken.

This establishes the regularity of the physical amplitude, but we must still connect it to the determinant representation from which the universal coefficients were obtained. We have already shown that the determinants reproduce the physical amplitude when $\operatorname{Re}\mu_a>3$. We have now also shown that the coefficients of the same physical amplitude admit a regular continuation near $\boldsymbol\mu=0$. Since the two representations agree in their initial convergence region, uniqueness of analytic continuation implies that the continued, fully symmetrized determinant coefficients must coincide with those of the physical amplitude.

This conclusion resolves the divergence problem identified at the beginning of this subsection. Although the individual mode and pump contributions in Eq.~\eqref{eq:app_chi_direct} can develop singularities when continued toward $\boldsymbol\mu=0$, their combination represents an amplitude whose perturbative coefficients are regular. These singularities must therefore cancel in the sum, leaving finite coefficients after the logarithmic derivatives in Eq.~\eqref{eq:app_chi_log_derivatives} are taken.

The only remaining step concerns the upper integration limit, which we have kept finite throughout the argument. Unlike the lower integration limit, the upper integration limit does not introduce the same difficulty. At every fixed SR order, the mode potential decreases at large $x$ as $x^{-2}$ multiplied by a finite polynomial in $\ln x$. The corresponding integrals converge with the outgoing prescription, allowing the upper cutoff to be removed.

We have therefore established that the determinant representation can be continued from its initial convergence region to the logarithmic sources entering Eq.~\eqref{eq:app_M_finite}. Although the separate mode and pump contributions need not be finite, their combination gives well-defined universal coefficients at every finite SR order.

\section{Derivation of the Arbitrary-Order Consistency Relations}
\label{app:Consistency_Relations}

In this appendix, we derive the explicit construction of the inflationary consistency relations presented in Sec.~\ref{sec:Relations} at arbitrary fixed SR order $W$. These relations connect the tensor spectral observables to the scalar spectrum and the tensor-to-scalar ratio without requiring knowledge of the underlying inflationary background. 

At leading order, the familiar single-field consistency relation $n_t=-r/8$ follows directly from $n_t=-2\epsilon_1$ and $r=16\epsilon_1$, where $n_t=d\ln\mathcal P_t/d\ln k$ is the tensor spectral index and $r=\mathcal P_t/\mathcal P_s$ is the tensor-to-scalar ratio. Taking successive derivatives with respect to $\ln k$ generates the well-known hierarchy of tensor consistency relations~\cite{Cortes:2006ap} in Eq.~\eqref{eq:intro_flow}. As shown in Sec.~\ref{sec:leading_order}, this entire LO hierarchy can be expressed in closed form using Bell polynomials. Our goal is to extend this construction beyond LO, obtaining explicit consistency relations at arbitrary fixed SR order without relying on recursive calculations.

Beyond leading order, the primordial spectra receive the corrections $M_A$ derived in Appendix~\ref{app:spectra_proof}, and the relation between the spectral observables and the inflationary background becomes more complicated. In particular, applying the LO reconstruction to the full spectra does not recover the exact background variables, but introduces higher-order corrections. To obtain consistency relations expressed entirely in terms of observables, we must therefore invert this corrected relation at arbitrary fixed SR order. Yet another difficulty is that the higher derivatives of the tensor spectrum are taken with respect to $\ln k$, while the inflationary background is naturally described in terms of the number of e-folds $N=\ln a$. The two derivatives are related by $d/d\ln k=hD_N$, where $D_N=d/dN$ and $h=-\eta aH$. Since $h$ evolves with the background, successive scale derivatives act on both the physical quantities and $h$ itself. We must therefore account for these contributions explicitly at arbitrary derivative order.

Our derivation proceeds in two steps. First, in Sec.~\ref{app:background_inverse}, we obtain a finite expression for the perturbative background inversion, allowing the inflationary parameters to be eliminated in favor of spectral observables. Second, in Sec.~\ref{app:closed_derivatives}, we derive a closed expression for the repeated scale derivatives entering the tensor spectral hierarchy. Combining these results provides the consistency relations at any fixed SR order without requiring recursive inversion of the background or repeated application of the scale-derivative operator.

\subsection{Perturbative Background Inversion}
\label{app:background_inverse}

We first derive the perturbative inversion required to express the consistency relations entirely in terms of spectral observables. As discussed in Sec.~\ref{sec:Background_removal}, applying the LO reconstruction to the full primordial spectra does not recover the exact background variables. Instead, it gives $\widetilde{\mathbf b}=\mathbf b+\delta\mathbf b_W(\mathbf b)$, where $\mathbf b$ denotes the true inflationary background and $\delta\mathbf b_W$ contains the corrections through SR order $W$. Each component of this shift starts one SR order above the corresponding background variable. Our goal is to invert this relation and express any background quantity $f(\mathbf b)$ directly in terms of the reconstructed variables.

We begin by explaining why this inversion can be performed at any fixed SR order. The forward shift defines an operator $\Delta_W$ acting on a background function as $\Delta_W f(\mathbf b)=f(\mathbf b+\delta\mathbf b_W(\mathbf b))-f(\mathbf b)$. Since every component of $\delta\mathbf b_W$ starts one SR order above the corresponding variable, applying $\Delta_W$ increases the SR order of $f$ by at least one. Consequently, if $f$ starts at SR order $d$, the successive applications $\Delta_W^n f$ contribute only from order $d+n$. At fixed order $W$, all terms with $n>W-d$ can therefore be neglected. This explains why the inverse in Eq.~\eqref{eq:finite_inverse} contains only finitely many terms.

Although this result establishes that the inversion is finite, evaluating it through repeated applications of $\Delta_W$ becomes increasingly expensive at higher orders. We therefore derive an equivalent expression involving only ordinary derivatives of the known forward shift. To make this explicit, we write $\mathbf b=(b_0,\ldots,b_W)$, with $b_0=\epsilon_1$ and $b_i=q_i\equiv D_N^i\ln\epsilon_1$ for $i\geq1$. We also denote the components of the shift by $\delta b_i\equiv(\delta\mathbf b_W)_i$ and introduce the multi-index $\mathbf m=(m_0,\ldots,m_W)$, with $|\mathbf m|=\sum_i m_i$, $\mathbf m!=\prod_i m_i!$, $\partial_{\mathbf b}^{\mathbf m}=\prod_i\partial_{b_i}^{m_i}$, and $(\delta\mathbf b_W)^{\mathbf m}=\prod_i(\delta b_i)^{m_i}$. These definitions allow us to express the contributions from arbitrary products of the background shifts and their derivatives in a compact form.

The starting point is the ordinary change of variables $\mathbf c=\mathbf b+\delta\mathbf b_W(\mathbf b)$. Its Jacobian is
\begin{equation}
\mathcal J_W(\mathbf b)
=
\det\left[
\delta_{ij}+\partial_{b_j}\delta b_i
\right].
\end{equation}
To understand how this Jacobian enters the inverse, consider an arbitrary smooth function $\varphi(\mathbf c)$ whose value and derivatives vanish at the integration boundaries. Changing variables from $\mathbf c$ to $\mathbf b$ gives
\begin{equation}
\int f(\mathbf b(\mathbf c))\varphi(\mathbf c)\,d\mathbf c
=
\int f(\mathbf b)\mathcal J_W(\mathbf b)
\varphi\!\left(\mathbf b+\delta\mathbf b_W(\mathbf b)\right)d\mathbf b.
\end{equation}
The Jacobian appears because the integration variables have been changed, while the argument of $\varphi$ contains the known forward shift. We now expand this argument formally in powers of the shift using the multivariate Taylor expansion, retaining only the terms required at the chosen SR order
\begin{equation}
\varphi\!\left(\mathbf b+\delta\mathbf b_W(\mathbf b)\right)
=
\sum_{\mathbf m\in\mathbb N_0^{W+1}}
\frac{(\delta\mathbf b_W)^{\mathbf m}}{\mathbf m!}
\partial_{\mathbf b}^{\mathbf m}\varphi(\mathbf b).
\end{equation}
Substituting the Taylor expansion into the previous change-of-variables identity, we obtain
\begin{equation}
\begin{aligned}
\int f(\mathbf b(\mathbf c))\varphi(\mathbf c)\,d\mathbf c
&=\int f(\mathbf b)\mathcal J_W(\mathbf b)\sum_{\mathbf m}\frac{(\delta\mathbf b_W)^{\mathbf m}}{\mathbf m!}\partial_{\mathbf b}^{\mathbf m}\varphi(\mathbf b)\,d\mathbf b\\
&=\sum_{\mathbf m}\frac{1}{\mathbf m!}\int f(\mathbf b)(\delta\mathbf b_W)^{\mathbf m}\mathcal J_W(\mathbf b)\partial_{\mathbf b}^{\mathbf m}\varphi(\mathbf b)\,d\mathbf b.
\end{aligned}
\end{equation}
We now integrate each term by parts to transfer the derivatives from $\varphi$ to the remaining factors. Since $\varphi$ and its derivatives vanish at the integration boundaries, no boundary terms appear. Each derivative introduces a minus sign, giving
\begin{equation}
\int f(\mathbf b(\mathbf c))\varphi(\mathbf c)\,d\mathbf c
=
\sum_{\mathbf m}\frac{(-1)^{|\mathbf m|}}{\mathbf m!}
\int\varphi(\mathbf b)\partial_{\mathbf b}^{\mathbf m}
\left[
f(\mathbf b)(\delta\mathbf b_W)^{\mathbf m}\mathcal J_W(\mathbf b)
\right]d\mathbf b.
\end{equation}
Since this equality holds for any $\varphi$, we can identify the expressions multiplying $\varphi$ on both sides after relabeling the integration variable. We therefore obtain the inverse directly as
\begin{equation}
\boxed{
E_W[f](\mathbf o)
=
\left.
\left[
\sum_{|\mathbf m|\leq W-d}
\frac{(-1)^{|\mathbf m|}}{\mathbf m!}
\partial_{\mathbf b}^{\mathbf m}
\left\{
f(\mathbf b)
(\delta\mathbf b_W)^{\mathbf m}
\mathcal J_W(\mathbf b)
\right\}
\right]_{\leq W}
\right|_{\mathbf b=\mathcal U_{\rm LO}^{-1}(\mathbf o)}.
}
\label{eq:app_inverse_closed}
\end{equation}
Here $\mathbf o$ denotes the spectral observables, and $\mathcal U_{\rm LO}^{-1}(\mathbf o)$ gives the background variables reconstructed at LO. The sum is restricted to $|\mathbf m|\leq W-d$ because higher-order contributions can be neglected at fixed SR order $W$. The formula therefore expresses the true background quantity $f$ entirely in terms of observables and the known forward shift, without explicitly solving for the inverse background map.

Finally, we explain why the sum in Eq.~\eqref{eq:app_inverse_closed} terminates and how the SR truncation must be performed. Since each factor $\delta b_i$ starts one order above $b_i$, the net contribution from each shift and its corresponding derivative increases the SR order by at least one. A term containing $|\mathbf m|$ shifts therefore starts at order $d+|\mathbf m|$, so only $|\mathbf m|\leq W-d$ can contribute through order $W$. However, the derivatives in Eq.~\eqref{eq:app_inverse_closed} can lower the SR order of the expressions on which they act. We must therefore perform these derivatives before truncating the result. Since $b_0=\epsilon_1$ starts at first SR order and $b_j=q_j$ at order $j$, a term indexed by $\mathbf m$ must be retained through intermediate order $W+m_0+\sum_{j=1}^{W}j m_j$ before differentiation. Only after all derivatives have been evaluated can the final expression be truncated consistently at order $W$. 	Note also that the full background reconstruction uses the observable vector $\mathbf o$ through $\alpha_{W-1}^s$ to determine $q_W$. However, every tensor relation contains a factor $r$, of SR weight one. Since $\alpha_j^s$ has weight $j+1$, only $j\leq W-2=n+p-1$ can occur in $F_{n,p}$.

Overall, Eq.~\eqref{eq:app_inverse_closed} provides an explicit, finite, and non-recursive expression for the perturbative background inverse at arbitrary fixed SR order. It depends only on the known forward shift and the LO reconstruction, without requiring the inverse to be constructed order by order. In principle, the remaining derivatives can be expanded into finite sums using multivariate Bell polynomials, yielding a fully algebraic closed-form expression. For simplicity, however, we maintain the derivative representation, which is more compact and equally suitable for evaluating the inverse at any desired order.

\subsection{Closed Scale-Derivative Identity}
\label{app:closed_derivatives}

We now derive the identity required to calculate the higher derivatives of the tensor spectrum with respect to $\ln k$. As discussed in Sec.~\ref{sec:Relations}, scale derivatives are related to derivatives with respect to the number of e-folds by $d/d\ln k=hD_N$, where $D_N=d/dN$. However, successive applications of this relation also differentiate $h$, which depends on the inflationary background. For example, $(hD_N)^2X=h^2D_N^2X+h(D_Nh)D_NX$. At higher orders, the number of terms increases, and evaluating them through repeated differentiation becomes increasingly cumbersome. Our goal is to express $(hD_N)^\nu X$ directly as a finite sum of ordinary derivatives of $X$, with coefficients determined explicitly by the background.

To derive this expression, we consider a displacement $u$ in $N$ and the corresponding displacement $t$ in $\ln k$, both measured from the same reference point. Since $d\ln k/dN=1/h$, these displacements satisfy $dt/du=h^{-1}(N+u)$. Integrating from $0$ to $u$, we obtain
\begin{equation}
t(u)=\int_0^u\frac{dv}{h(N+v)}.
\end{equation}
We now expand the integrand around $v=0$, with all background quantities evaluated at the reference point $N$:
\begin{equation}
\frac{1}{h(N+v)}
=
\sum_{j\geq0}\frac{v^j}{j!}D_N^j\left(\frac{1}{h}\right).
\end{equation}
Integrating each term and collecting the common factor $u/h$, we find
\begin{equation}
t(u)
=\sum_{j\geq0}\frac{u^{j+1}}{(j+1)!}D_N^j\left(\frac{1}{h}\right)
=\frac{u}{h}\left[1+\sum_{j\geq1}\frac{h}{j+1}D_N^j\left(\frac{1}{h}\right)\frac{u^j}{j!}\right].
\end{equation}
We can therefore write $t=u\mathcal A(u)/h$, with
\begin{equation}
\mathcal A(u)=1+\sum_{j\geq1}\gamma_j\frac{u^j}{j!},
\end{equation}
where the coefficients are identified directly from the previous expansion:
\begin{equation}
\gamma_j
=
\frac{h}{j+1}D_N^j\left(\frac{1}{h}\right)
=
\frac{1}{j+1}
\sum_{a=1}^{j}
(-1)^a \, a! \, h^{-a} \,
B_{j,a}(D_Nh,\ldots,D_N^{j-a+1}h).
\label{eq:app_gamma_closed}
\end{equation}
The second equality follows from the Bell-polynomial expansion of the derivatives of $h^{-1}$, expressing the coefficients $\gamma_j$ directly in terms of $h$ and its derivatives. These coefficients account for the evolution of $h$ along the inflationary background. For example, $\gamma_1=-D_Nh/(2h)$, while higher coefficients involve successive derivatives of $h$.

The relation between $t$ and $u$ allows us to calculate scale derivatives without applying $hD_N$ repeatedly. Indeed, expanding $X[N+u(t)]$ in powers of $t$ gives the successive scale derivatives of $X$, since the coefficient of $t^\nu$ is $(hD_N)^\nu X/\nu!$. We can therefore determine these derivatives by expressing $u$ as a function of $t$ and extracting the corresponding Taylor coefficients.

From the relation $t=u\mathcal A(u)/h$, Lagrange inversion gives
\begin{equation}
[t^\nu]X[N+u(t)]
=
\frac{h^\nu}{\nu}
[u^{\nu-1}]
D_NX(N+u)\mathcal A(u)^{-\nu},
\label{eq:app_clock_coefficient}
\end{equation}
where the brackets select the indicated Taylor coefficient. The important point is that the right-hand side depends only on ordinary derivatives with respect to $N$ and on the known coefficients $\gamma_j$. We can therefore evaluate it directly at any derivative order.

To make this explicit, we expand the two factors on the right-hand side of Eq.~\eqref{eq:app_clock_coefficient}. The first is the ordinary Taylor expansion $D_NX(N+u)=\sum_{k\geq1}u^{k-1}D_N^kX/(k-1)!$, while the second follows from the expansion of $\mathcal A(u)^{-\nu}$ in terms of partial Bell polynomials:
\begin{equation}
\mathcal A(u)^{-\nu}
=
\sum_{r\geq0}\frac{u^r}{r!}
\sum_{m=0}^{r}(-1)^m\nu^{\overline m}
B_{r,m}(\gamma_1,\ldots,\gamma_r).
\end{equation}
Here $\nu^{\overline m}=\nu(\nu+1)\cdots(\nu+m-1)$ is the rising factorial, with $\nu^{\overline 0}=1$. We adopt the conventions $B_{0,0}=1$ and $B_{r,0}=0$ for $r>0$. The partial Bell polynomials collect all products of the coefficients $\gamma_j$ that contribute to a given power of $u$.

We can now substitute these expansions into Eq.~\eqref{eq:app_clock_coefficient}. To obtain the coefficient of $u^{\nu-1}$, a term containing $D_N^kX$ must be multiplied by the term of order $u^{\nu-k}$ in $\mathcal A(u)^{-\nu}$. Consequently, only $1\leq k\leq\nu$ contributes, and the sum over $m$ terminates at $m=\nu-k$. Multiplying the resulting coefficient by $\nu!$, we obtain
\begin{equation}
(hD_N)^\nu X
=
\sum_{k=1}^{\nu}c_{\nu,k}D_N^kX,
\qquad
c_{\nu,k}
=
\binom{\nu-1}{k-1}h^\nu
\sum_{m=0}^{\nu-k}
(-1)^m\nu^{\overline m}
B_{\nu-k,m}(\gamma_1,\ldots,\gamma_{\nu-k}).
\label{eq:app_operator_closed}
\end{equation}
For $\nu=0$, the operator reduces to the identity. Equation~\eqref{eq:app_operator_closed} reproduces the second-order example given above, with $c_{2,2}=h^2$ and $c_{2,1}=hD_Nh$. More generally, it determines every coefficient directly from $h$ and its derivatives, without requiring the corresponding operators at lower derivative orders.

Finally, we determine which terms contribute at a given SR order. From Eq.~\eqref{eq:app_gamma_closed}, $\gamma_j$ starts at SR order $j+1$. Since the Bell polynomial $B_{r,m}$ contains products of $m$ coefficients whose indices sum to $r$, each such term starts at SR order $r+m$. This allows us to truncate the sum in Eq.~\eqref{eq:app_operator_closed} consistently. In particular, for $W=n+1+p$, the background term in Eq.~\eqref{eq:tensor_closed} receives contributions only from $m\leq p$, while the $M_t$ term receives contributions only from $m\leq p-1$. Together with $D_N^kq_j=q_{j+k}$ and $D_N^k\epsilon_1^a=\epsilon_1^aB_k(aq_1,\ldots,aq_k)$, this makes all derivatives entering the tensor consistency relations explicit at fixed $W$.

Equation~\eqref{eq:app_operator_closed} therefore provides a closed, finite, and non-recursive expression for scale derivatives at arbitrary order. Combined with the perturbative background inversion derived in Sec.~\ref{app:background_inverse}, it completes the construction required to express the higher-order tensor consistency relations entirely in terms of spectral observables.

\section{Extension to $k$-Inflation}
\label{app:noncanonical}

The derivation of the primordial spectra at arbitrary SR order and the corresponding consistency relations presented in the previous sections have assumed canonical single-field inflation. In this appendix, we extend the construction to single-field $k$-inflation~\cite{Garriga:1999vw} with a slowly varying scalar sound speed $c_s$.

The main difference is that scalar perturbations propagate with sound speed $c_s$, while tensor perturbations maintain their canonical propagation speed $c_t^2=1$. Consequently, the scalar mode equation and the relation between the scalar and tensor reference scales must be modified. We show that these modifications leave the universal coefficients entering the primordial spectra unchanged and allow the consistency relations to be constructed at arbitrary fixed SR order.

To describe the sound-speed evolution, we introduce the background and spectral hierarchies
\begin{equation}
s_j\equiv D_N^j\ln c_s,\qquad
\alpha_n^{c_s}\equiv\frac{d^{n+1}\ln c_s}{d(\ln k)^{n+1}}
=(hD_N)^{n+1}\ln c_s.
\label{eq:cs_inputs}
\end{equation}
Here $s_j=\mathcal O(\mathrm{SR}^j)$ for $j\geq1$, while $c_s$ and $\ln c_s$ are counted as zeroth order. The spectral hierarchy is defined using the tensor scale derivative $hD_N$, since both spectra will eventually be compared at a common tensor reference scale. The two hierarchies are related by the closed scale-derivative identity derived in Sec.~\ref{app:closed_derivatives}, giving
\begin{equation}
\alpha_n^{c_s}=\sum_{k=1}^{n+1}c_{n+1,k}(h)s_k,
\label{eq:alpha_cs}
\end{equation}
where the coefficients $c_{n+1,k}$ are fixed by Eq.~\eqref{eq:app_operator_closed}.

With these definitions in place, we proceed with the generalization in three steps. First, in Sec.~\ref{app:cs_spectra}, we extend the derivation of the primordial spectra to $k$-inflation, including their evaluation at a common reference scale. Second, in Sec.~\ref{app:cs_background}, we generalize the perturbative background elimination to include the sound-speed hierarchy. Finally, in Sec.~\ref{app:cs_relations}, we extend the construction of the consistency relations to arbitrary fixed SR order.

\subsection{Generalization of Primordial Spectra at Arbitrary Order}
\label{app:cs_spectra}
We begin by showing how the scalar mode equation can be brought into the same form used in the canonical construction. In $k$-inflation, the scalar Mukhanov--Sasaki equation contains a propagation term $c_s^2k^2$, with the canonical variable $v_{s,k}=z_s\mathcal R_k$ and $z_s=aM_{\rm Pl}\sqrt{2\epsilon_1}/c_s$. Introducing the acoustic time $d\tau=c_s\,d\eta$ and the rescaled variable $w_{s,k}=\sqrt{c_s}\,v_{s,k}$, we obtain
\begin{equation}
\partial_\tau^2w_{s,k}
+
\left(
k^2-\frac{\partial_\tau^2\widetilde z_s}{\widetilde z_s}
\right)w_{s,k}=0,
\end{equation}
where $\widetilde z_s=\sqrt{c_s}\,z_s=aM_{\rm Pl}\sqrt{2\epsilon_1/c_s}$. Thus, the scalar perturbations satisfy the same type of mode equation as in the canonical case, but with acoustic time and a modified pump field.

We now introduce the scalar variable $x_s=-k\tau$ and the corresponding clock $h_s=-\tau aH/c_s$. Differentiating its definition with respect to $N$ gives $D_Nh_s=h_s(1-\epsilon_1-s_1)-1$. This has the same structure as the canonical clock equation, with $\epsilon_1$ replaced by $\epsilon_1+s_1$. Therefore, the explicit SR expansion of $h_s$ follows from Eq.~\eqref{eq:app_h_closed} through this replacement, including the corresponding background derivatives.

The tensor clock and source remain unchanged, since tensor perturbations are unaffected by the scalar sound speed. Instead, the scalar source can be obtained from the modified pump field. Defining $Q_s\equiv D_N\ln\widetilde z_s=1+(q_1-s_1)/2$ and using $d/d\tau=(aH/c_s)D_N$, we find
\begin{equation}
g_s=h_s^2\left[
D_NQ_s+Q_s^2+(1-\epsilon_1-s_1)Q_s
\right]-2.
\label{eq:cs_source}
\end{equation}

The important consequence is that both mode equations maintain the same free de Sitter operator used in Appendix~\ref{app:spectra_proof}. The sound-speed dependence is entirely contained in the scalar source and its SR expansion. Therefore, the universal coefficients $\chi_{m;\mathbf j}$ in Eq.~\eqref{eq:app_M_finite} remain unchanged. The scalar correction $M_s$ follows from the same construction\footnote{The tensor correction $M_t$ maintains its canonical expression.}, replacing $h$ with $h_s$ and using
\begin{equation}
g_{s,j}=\frac{1}{j!}[(-h_sD_N)^jg_s]_{\leq W}.
\end{equation}
At their respective reference scales, defined by $k_s(N)=-1/\tau(N)$ and $k_t(N)=-1/\eta(N)$, the resulting spectra are
\begin{equation}
\mathcal P_s(k_s(N))
=\frac{H^2}{8\pi^2M_{\rm Pl}^2\epsilon_1c_s}e^{M_s},
\qquad
\mathcal P_t(k_t(N))
=\frac{2H^2}{\pi^2M_{\rm Pl}^2}e^{M_t}.
\label{eq:cs_spectra}
\end{equation}
These expressions have the same structure as their canonical counterparts, with the additional factor $c_s$ in the scalar normalization and the modified correction $M_s$.

However, the scalar and tensor reference scales generally differ at the same background time. Since the tensor-to-scalar ratio and the consistency relations compare both spectra at the same wavenumber, we must account for this difference. We choose the common pivot $k_*=k_t(N_*)$ and define the logarithmic separation between the two reference scales as
\begin{equation}
L\equiv\ln\frac{k_t(N)}{k_s(N)}
=\ln c_s+\ln h_s-\ln h.
\label{eq:cs_pivot_separation}
\end{equation}
The second equality follows directly from the definitions of $h$, $h_s$, and the two reference scales.

To express the scalar spectrum at the tensor reference scale, we expand $\ln\mathcal P_s$ around $k_s(N)$ in powers of the logarithmic displacement $L$. Since $d\ln k_s/dN=1/h_s$, derivatives with respect to the scalar reference scale are generated by $h_sD_N$. We therefore obtain
\begin{equation}
\ln\mathcal P_s(k_t(N))
=
\left[
\ln\mathcal P_s(k_s(N))
+\sum_{j=1}^{W}\frac{L^j}{j!}
(h_sD_N)^j\ln\mathcal P_s(k_s(N))
\right]_{\leq W}.
\label{eq:cs_pivot_spectrum}
\end{equation}
In Eq.~\eqref{eq:cs_pivot_spectrum}, the Taylor expansion is performed at fixed background time $N$. The separation $L$ is therefore held fixed, and the derivatives act only on the scalar spectrum. However, when we subsequently calculate the spectral derivatives, we must account for the fact that $L$ also evolves with $N$. From Eq.~\eqref{eq:cs_pivot_separation}, we find $D_NL=1/h-1/h_s$.

After this translation, the scalar and tensor spectra are both expressed at $k_t(N)$. Their derivatives with respect to the common wavenumber are therefore generated by $hD_N$, including the contributions from the evolution of $L$. The corresponding spectral hierarchies can then be calculated using the closed scale-derivative identity derived in Sec.~\ref{app:closed_derivatives}.

Finally, since $c_s$ is not required to be close to unity, the separation $L$ need not be small. Equation~\eqref{eq:cs_pivot_spectrum} is therefore not an expansion in small $L$, but an expansion in SR order, since successive scale derivatives introduce higher-order SR contributions. The truncation at order $W$ remains valid provided these contributions, including those enhanced by powers of $\ln c_s$, remain perturbatively controlled. All sound-speed quantities are evaluated at the common background time $N_*$.

\subsection{Generalization of Background Elimination}
\label{app:cs_background}

Having obtained the primordial spectra at a common reference scale, we now extend the background-elimination procedure developed for canonical inflation. 

The first step is to determine how the LO reconstruction changes when the scalar sound speed is allowed to evolve. At LO, $r=16\epsilon_1c_s$ and $n_t=-2\epsilon_1$, while the scalar spectral index satisfies $\alpha_0^s=-2\epsilon_1-q_1-s_1$. The sound-speed evolution therefore enters the scalar tilt through the additional contribution $s_1$, whereas the tensor tilt maintains its canonical dependence on $\epsilon_1$. These relations give $n_t=-r/(8c_{s})$ and $q_1=n_t-\alpha_0^s-\alpha_0^{c_s}$ at LO. Taking successive scale derivatives extends the same structure to the complete LO hierarchy. In particular, the canonical closed expression $F_{n,0}$ remains applicable after the replacements $r\longrightarrow r/c_{s}$ and $\alpha_j^s\longrightarrow\alpha_j^s+\alpha_j^{c_s}$. Denoting the resulting function by $F_{n,0}^{(c_s)}$, we obtain the LO reconstruction of the background variables:
\begin{equation}
(\epsilon_1)_{\rm LO}=\frac{r}{16c_{s}},\qquad
(s_j)_{\rm LO}=\alpha_{j-1}^{c_s},\qquad
(q_j)_{\rm LO}=F_{j-1,0}^{(c_s)}
-\alpha_{j-1}^s-\alpha_{j-1}^{c_s}.
\label{eq:cs_LO_inverse}
\end{equation}

Beyond LO, the reconstructed variables receive higher-order corrections from the primordial spectra and from the difference between the scalar and tensor reference scales. To eliminate these corrections, we apply the perturbative inversion derived in Sec.~\ref{app:background_inverse} to the enlarged background $\mathbf b_{c_s}=(\epsilon_1,q_1,\ldots,q_W,s_1,\ldots,s_W)$. Applying the LO reconstruction in Eq.~\eqref{eq:cs_LO_inverse} to the full forward map produces a perturbative shift in these variables. As in the canonical case, each component of the shift starts one SR order above the corresponding background variable. The finite inversion therefore preserves the same structure as Eq.~\eqref{eq:app_inverse_closed}, with the multi-index and Jacobian extended to include the sound-speed variables. The pivot value $c_{s*}$ is treated as an additional fixed input, while its evolution is encoded in the hierarchy $s_j$. No new inversion procedure is required.

We can therefore define the resulting background-elimination operator $E_W^{(c_s)}$. Like its canonical counterpart, it provides a finite, explicit, and non-recursive expression for any background quantity at the desired SR order, now in terms of the enlarged set of scalar and sound-speed observables.

\subsection{Generalization of Consistency Relations at Arbitrary Order}
\label{app:cs_relations}

We can now construct the tensor consistency relations by combining the generalized primordial spectra derived in Sec.~\ref{app:cs_spectra} with the enlarged background inversion obtained above. Since both spectra have been expressed at the same reference scale, their spectral derivatives can be calculated using the common operator $hD_N$.

The construction proceeds exactly as in Sec.~\ref{sec:Relations}, replacing $E_W$ with $E_W^{(c_s)}$ in the background-elimination step. The tensor expression and its closed scale-derivative coefficients $c_{n+1,k}(h)$ remain unchanged. All additional contributions arise from the modified scalar spectrum, the separation between the scalar and tensor reference scales, and the enlarged background reconstruction.

For $W=n+1+p$, the resulting relations take the form $\alpha_n^t=F_{n,p}^{(c_s)}+\mathcal O(\mathrm{SR}^{W+1})$, where the full background inversion uses the enlarged observable vector $\mathbf o_{c_s}=(r,\alpha_0^s,\ldots,\alpha_{W-1}^s,c_{s},\alpha_0^{c_s},\ldots,\alpha_{W-1}^{c_s})$. The final function $F_{n,p}^{(c_s)}$ is independent of $\alpha_{W-1}^s$ and $\alpha_{W-1}^{c_s}$. Its scalar and sound-speed derivative arguments extend only through $W-2=n+p-1$. Both derivative hierarchies are absent when $n=p=0$, leaving $F_{0,0}^{(c_s)}=-r/(8c_s)$.

The consistency hierarchy therefore remains explicit and non-recursive at arbitrary $n$ and finite $p$, provided the sound-speed hierarchy is supplied alongside the scalar observables. In the limit $c_s=1$ and $s_j=0$, the construction reduces to the canonical consistency relations.

\section{Implementation, Validation and Comparison with Known Results}
\label{app:validation}

In this appendix, we first describe the implementation of our formalism in a symbolic algorithm that, for any pair $(n,p)$, consistently generates the primordial spectra and the corresponding consistency relations at the desired SR order. We then explain how the different ingredients of the construction are validated, including comparisons between independent calculations and checks of the arbitrary-order expressions.

We subsequently compare our results with the existing literature, showing that the construction reproduces the known consistency relations at LO, NLO, N2LO, and N3LO, before extending the validation through N10LO. These comparisons provide complementary tests of the primordial spectra, the perturbative background inversion, and the resulting consistency relations, for both canonical single-field inflation and $k$-inflation. For a survey of representative studies of the consistency-relation hierarchy and its applications to inflationary observables, primordial gravitational waves, and parameter inference, see, e.g., Refs.~\cite{Liddle:1994cr,Turner:1995ge,Lidsey:1995np,Hui:2001ce,Bartolo:2001rt,Kinney:2002qn,Ashoorioon:2005ep,Gong:2007ha,Cheng:2012je,Palma:2014faa,Boyle:2014kba,Cabass:2015jwe,Huang:2015gca,Meerburg:2015zua,Brooker:2016imi,Broy:2016zik,Giare:2019snj,Cano:2020oaa,Capurri:2020qgz,Giare:2020plo,Giare:2020vhn,Giare:2020vss,Vagnozzi:2020gtf,Forconi:2021que,Kinney:2021nje,Li:2021nqa,Benetti:2021uea,Giare:2022wxq,Giovannini:2022etg,German:2023euc,Giare:2023wzl,Bianchi:2024qyp,Cecchini:2024xoq,Giare:2024sdl,Balkenhol:2025wms,Brax:2025osk,Haddad:2026hsf,Giare:2026oti} and references therein.

The symbolic implementation of our formalism is publicly available at \faGithub\ \url{https://github.com/williamgiare/slow-roll-inflation}. The repository also contains the scripts required to reproduce the validation tests presented in this appendix. In addition, for both canonical inflation and $k$-inflation, we provide the explicit expressions for the scalar and tensor primordial spectra from LO through N10LO ($p=0,\ldots,10$), together with the corresponding consistency relations for tensor spectral derivatives up to $n=3$.

The importance of an explicit arbitrary-order construction becomes particularly evident at high SR orders, where the number of contributions and the complexity of the resulting expressions grow dramatically. At N10LO, the fully expanded expressions for the primordial spectra occupy approximately 75 pages for canonical inflation and more than 2200 pages for $k$-inflation. Reproducing such expressions in the manuscript would therefore be impractical, and we instead make them available in the GitHub repository. This rapid growth highlights the value of the general formulas derived in this work, which provide a compact and systematic representation of the primordial spectra and consistency relations at arbitrary SR order.

\subsection{Implementation and Validation}

\subsubsection{Implementation}
We implement the construction using symbolic polynomial algebra. All calculations are organized according to the SR expansion. In particular, each background variable is assigned its SR weight, namely $w(\epsilon_1)=1$, $w(q_j)=w(s_j)=j$, and $w(\ln c_s)=0$. These weights determine which contributions must be preserved at a given order. 

To specify the overall order, we input the relative SR order $p$ of the primordial spectra and, when constructing a consistency relation, the tensor derivative order $n$. The pair $(n,p)$ determines the absolute SR order $W=n+1+p$. 

After the order is specified, products and substitutions are truncated accordingly, while derivatives are evaluated with particular care as they can lower the SR weight of an expression. As discussed in Appendix~\ref{app:background_inverse}, this requires preserving sufficiently high intermediate orders before performing the final truncation.

The first step is to compute the universal coefficients in the primordial spectra. We use the fully symmetrized coefficients defined in Eq.~\eqref{eq:app_chi_direct}, with the corresponding symmetric auxiliary function in Eq.~\eqref{eq:app_C_auxiliary}. Since they are independent of the inflationary background, they can be determined separately from the remaining calculation. Starting from the source expansion in Eq.~\eqref{eq:g_expansion}, where each coefficient $g_{A,j}$ carries SR weight $j+1$, we implement two complementary methods to obtain these coefficients.
\begin{itemize}
\item The first method determines the coefficients numerically. In particular, we expand the mode and pump equations~\eqref{eq:app_mode_pump_equations} in powers of the source amplitudes and solve the resulting equations with the Bunch--Davies condition~\eqref{eq:app_BD_condition} for the mode and the normalized growing solution for the pump. The coefficients of $\ln|J_A|^2$ are then extracted from the super-horizon amplitude defined in Eq.~\eqref{eq:app_J}. Numerical convergence is tested by varying the initial conditions, integration limits, tolerances, and working precision. Once determined, the coefficients are represented by fixed rational approximations, allowing the subsequent symbolic calculations to proceed using exact rational arithmetic.
\item The second method evaluates the coefficients directly from the analytic expressions derived in Appendix~\ref{app:spectra_proof}. Starting from the free de Sitter kernels, we calculate the ordered pump and mode integrals in Eqs.~\eqref{eq:app_ordered_pump} and~\eqref{eq:app_ordered_mode}, which determine the auxiliary function $\mathcal C_m$ in Eq.~\eqref{eq:app_C_auxiliary}. The universal coefficients are then extracted through the derivatives in Eq.~\eqref{eq:app_chi_log_derivatives}, combining the mode and pump contributions according to the continuation established in Sec.~\ref{sec:app_convergence}. This procedure evaluates the general representation in Eq.~\eqref{eq:app_chi_direct}, maintaining exact analytic expressions without constructing perturbatively corrected mode functions.
\end{itemize}
These two calculations provide independent determinations of the same universal coefficients, whose agreement is tested through N10LO in the following subsection. Neither method uses previously published spectral coefficients as input. Moreover, the coefficients depend only on the SR order of the spectra: increasing the tensor derivative order $n$ at fixed $p$ requires additional background derivatives, but no new universal coefficients.

Once the universal coefficients are known, the remaining task is to construct the spectra and eliminate their dependence on the inflationary background. For this stage of the calculation, we also implement two complementary approaches.
\begin{itemize}
\item The first approach is iterative. The clock $h$, the source coefficients, and the spectral derivatives are generated through successive applications of the corresponding background operators. The resulting observables are then used to eliminate the inflationary parameters through the finite difference expansion in Eq.~\eqref{eq:EW}.

\item The second approach uses the closed-form expressions derived in the previous appendices. In particular, we evaluate the Bell-polynomial expression for $h$ in Eq.~\eqref{eq:app_h_closed}, the closed scale-derivative identity in Eq.~\eqref{eq:app_operator_closed}, and the perturbative background inverse in Eq.~\eqref{eq:app_inverse_closed}. The ordinary derivatives appearing in these expressions are evaluated using the background Taylor expansions, preserving the intermediate SR orders required before the final truncation.
\end{itemize}
Both implementations construct the spectral corrections $M_A$ using Eq.~\eqref{eq:app_M_finite}, and neither requires lower-order tensor consistency relations as input. In practice, we use the iterative implementation because it avoids generating large intermediate expressions containing terms that subsequently cancel. The closed implementation, however, provides an independent evaluation of the general formulas, allowing their algebraic equivalence to be tested directly. Since both implementations use the same universal coefficients, their accuracy is established separately through the analytic and numerical comparisons described below.

The extension to $k$-inflation follows the same procedure, with the modifications derived in Appendix~\ref{app:noncanonical}. The scalar clock and source are replaced by their acoustic counterparts, while the tensor calculation remains unchanged. Since the scalar and tensor reference scales differ, we first translate the scalar spectrum to the tensor reference scale using Eq.~\eqref{eq:cs_pivot_spectrum}. We then construct the spectral observables and eliminate the enlarged set of background variables through the corresponding perturbative inversion. In this step, the sound-speed parameters $s_j$ are expressed in terms of the spectral quantities $\alpha_{j-1}^{c_s}$ using Eq.~\eqref{eq:alpha_cs}. The resulting consistency relations therefore depend on $c_s$ and its spectral hierarchy, rather than on the background derivatives $s_j$, with the sound-speed evolution consistently accounted for in the reference-scale translation.

\subsubsection{Validation}

Our validation is designed to test the different ingredients of the construction independently. We first examine the universal coefficients through analytic calculations and independent numerical solutions of the mode equations. We then test the resulting primordial spectra and compare the iterative and closed implementations to verify their algebraic equivalence. Finally, we check that the consistency relations correctly eliminate the inflationary background.

We begin with the universal coefficients, which require a separate validation because both implementations of the background algebra use the same coefficient values. We perform three complementary checks:
\begin{itemize}
\item We first test the determinant representation in Eq.~\eqref{eq:app_J_determinants} against direct numerical solutions of the mode and growing-pump equations, using compact sources of either sign. Refining the quadrature from $24$ to $48$ and $96$ points per interval produces convergence to the same complex amplitude, with final relative differences below $3\times10^{-9}$. This provides a direct check of the determinant construction independently of the subsequent symbolic calculations.

\item We then examine the analytic representation of the universal coefficients. The construction is first checked against the exact constant-source Hankel solution, the one-insertion Gamma expression, and the two-insertion kernel. We extend the comparison to all $138$ independent, fully symmetrized universal coefficients required through N10LO, including all $67$ mixed-source combinations. Each coefficient is evaluated from the analytic expressions derived in Appendix~\ref{app:spectra_proof} and compared individually with its numerical determination. The auxiliary poles introduced by the analytic representation cancel identically, and every coefficient agrees with the numerical result within its independently estimated uncertainty. This comparison also verifies the analytic construction at higher insertion orders, beyond the cases for which simple closed expressions are available.

\item As a third independent check, we use a separate solver for the linear mode and growing-pump equations to verify the complete coefficient set. Its initial conditions, integration range, and numerical precision are refined independently of the other calculations. Together, these checks test the explicit analytic representation of the universal coefficients, as well as the normalization and source dependence entering the primordial spectra.
\end{itemize}

Having validated the universal coefficients, we examine the resulting primordial spectra through two additional tests:
\begin{itemize}
\item We compare the reconstructed scalar and tensor spectra with independent numerical integrations of the corresponding mode equations on nonconstant inflationary backgrounds. By rescaling the background parameters, we examine how the difference between the truncated SR expressions and the numerical solutions decreases as the background approaches the slow-roll limit. This allows us to test the expected order of the omitted contributions. We further vary the initial integration time and the final super-horizon evaluation time to distinguish errors associated with the SR truncation from those introduced by the numerical integration.

\item For $k$-inflation, the scalar calculation is additionally checked through an independent effective-pump construction, which provides a separate derivation of the acoustic source and its normalization. Exact solutions with evolving sound speed are also used to verify the translation between scalar and tensor reference scales. These comparisons are particularly important because the evolution of $c_s$ introduces contributions that cannot be tested in the constant-sound-speed limit.
\end{itemize}

We finally turn to the algebraic construction of the consistency relations. In this case, we perform three complementary checks:
\begin{itemize}
\item We first compare the iterative and closed implementations at successive stages of the calculation, including the clock $h$, the source coefficients, the corrections $M_A$, the primordial spectra, the forward map from background variables to observables, and the final consistency relations. In canonical inflation, the complete comparison covers $p=0,\ldots,4$ and $n=0,\ldots,3$, with the LO hierarchy additionally checked through $n=8$. For $k$-inflation, the corresponding comparison covers $p=0,\ldots,3$ and the same derivative orders. In all cases, the resulting polynomials agree exactly when evaluated using the same fixed numerical representations of the universal coefficients.

\item Although this agreement verifies the algebraic equivalence of the two implementations, it does not prove that the observable reconstruction correctly eliminates the inflationary background. We therefore substitute the forward observable map into each generated consistency relation and require that it reproduces the tensor observable calculated directly from the background:
\begin{equation}
\left[
F_{n,p}\!\left(\mathcal U_W(\mathbf b)\right)
-\alpha_n^t(\mathbf b)
\right]_{\leq W}=0,
\label{eq:app_validation_forward}
\end{equation}
where $W=n+1+p$. Since this identity is evaluated as a polynomial in the general background variables, it tests the cancellation of the background dependence without restricting the variables to particular numerical values. We also verify that every term has the correct SR weight, that the final expressions contain only the prescribed observables, and that the tensor relations vanish regularly when $r=0$. The forward-substitution test covers all $64$ consistency relations provided in the repository, comprising $32$ canonical and $32$ $k$-inflation relations. In every case, the residual vanishes identically through the required SR order. These substitutions use exact rational arithmetic, with the numerically determined universal coefficients held fixed at their converged values.

\item We perform further checks of the observable reconstruction in both inflationary scenarios. For canonical inflation, we compare the results with a separate construction based on the observable flow and a one-variable inversion, finding agreement through N3LO. For $k$-inflation, the forward-substitution test uses the enlarged observable and background maps. We additionally verify both directions of the conversion between $s_j$ and $\alpha_{j-1}^{c_s}$ through $j=10$, together with the recovery of the canonical limit when the sound-speed contributions are removed.
\end{itemize}

These comparisons independently test the universal spectral coefficients and the algebraic construction of the consistency relations, including the elimination of the inflationary background. We now compare the resulting consistency relations with the existing literature, starting at LO and extending to higher SR orders.

\subsection{LO Results}

\subsubsection{Canonical Inflation}
We first consider the leading-order hierarchy. Setting $p=0$ gives $W=n+1$. At this order, the corrections contained in $M_t$ contribute only beyond the required SR accuracy, while the scale-derivative coefficients in Eq.~\eqref{eq:app_operator_closed} retain only the $m=0$ contribution. Since $B_{n+1-k,0}$ vanishes unless $k=n+1$, the general construction reduces to
\begin{equation}
F_{n,0}
=
-2\left.
D_N^n\epsilon_1
\right|_{\mathbf b=\mathcal U_{\rm LO}^{-1}(\mathbf o)}.
\label{eq:app_LO_reduction}
\end{equation}
The corrections introduced by the perturbative inverse also start beyond order $n+1$ and can therefore be neglected.

We can recover the closed LO hierarchy derived in the main text directly from this expression. Introducing $L=\ln(k/k_\star)$, the LO relation between scale and background derivatives gives $D_N=d/dL$. Using $\epsilon_1(0)=r/16$ and defining
\begin{equation}
K_0=1,
\qquad
K_j=B_j(-\alpha_0^s,\ldots,-\alpha_{j-1}^s),
\end{equation}
the scalar spectral evolution yields
\begin{equation}
-2\epsilon_1(L)
=
-\frac{d}{dL}
\ln\left[
1+\frac{r}{8}
\sum_{j=0}^{n}
K_j\frac{L^{j+1}}{(j+1)!}
\right]
+\mathcal O(L^{n+1}).
\end{equation}
Taking $n$ derivatives at $L=0$, we obtain
\begin{equation}
F_{n,0}
=
\sum_{q=1}^{n+1}
\left(-\frac{r}{8}\right)^q
(q-1)!
B_{n+1,q}
(K_0,\ldots,K_{n+1-q}).
\label{eq:app_LO_recovery}
\end{equation}
This is precisely the closed-form expression in Eq.~\eqref{eq:LO_closed_form}. It therefore reproduces the complete LO hierarchy conventionally generated through the recursive relations in Eq.~\eqref{eq:intro_flow}.

To illustrate this equivalence explicitly, we evaluate the first three members of the hierarchy. Writing $x=r/8$, the required coefficients are
\begin{equation}
K_0=1,\qquad
K_1=-\alpha_0^s,\qquad
K_2=(\alpha_0^s)^2-\alpha_1^s.
\end{equation}
For $n=0$, only the contribution $B_{1,1}(K_0)=K_0$ appears, immediately giving the standard tensor consistency relation
\begin{equation}
n_t=-\frac{r}{8}.
\end{equation}

For $n=1$, the relevant Bell polynomials are $B_{2,1}=K_1$ and $B_{2,2}=K_0^2$. Substituting them into Eq.~\eqref{eq:app_LO_recovery}, we find $\alpha_1^t=-xK_1+x^2K_0^2=x(\alpha_0^s+x)$. Identifying $\alpha_0^s=n_s-1$ and $\alpha_1^t=\alpha_t$, this becomes
\begin{equation}
\alpha_t
=
\frac{r}{8}
\left(
n_s-1+\frac{r}{8}
\right).
\end{equation}

The next member, $n=2$, involves $B_{3,1}=K_2$, $B_{3,2}=3K_0K_1$, and $B_{3,3}=K_0^3$. Their substitution gives $\alpha_2^t=x[\alpha_1^s-(\alpha_0^s)^2-3x\alpha_0^s-2x^2]$. Using $\alpha_1^s=\alpha_s$ and $\alpha_2^t=\beta_t$, we recover
\begin{equation}
\beta_t
=
\frac{r}{8}
\left[
\alpha_s
-(n_s-1)^2
-\frac{3r}{8}(n_s-1)
-2\left(\frac{r}{8}\right)^2
\right].
\end{equation}

These are the first three consistency relations of the LO hierarchy in Ref.~\cite{Cortes:2006ap}. The agreement is obtained by evaluating the closed expression directly, without using the preceding tensor relations as input. Higher members follow in the same way, with $\alpha_3^t$ providing the first contribution involving the scalar running of the running, $\alpha_2^s=\beta_s$.

We have also tested the equivalence between the closed-form expression and the recursive LO hierarchy at substantially higher derivative orders. Direct symbolic comparison establishes polynomial equality through $n=8$. To extend this validation, we evaluate the two constructions independently using exact rational arithmetic, finding agreement through $n=1000$. We further confirm this agreement using finite-field calculations over three prime fields, with three independent scalar input sequences for each field. These complementary checks extend the validation to derivative orders far beyond those accessible through direct symbolic polynomial comparison.

\subsubsection{$k$-Inflation}

The LO tensor expression remains unchanged in $k$-inflation, but the background reconstruction must include the sound-speed hierarchy. Applying Eq.~\eqref{eq:cs_LO_inverse} to the canonical closed expression gives
\begin{equation}
F_{n,0}^{(c_s)}
=
F_{n,0}\!\left(
\frac{r}{c_s},
\alpha_0^s+\alpha_0^{c_s},
\ldots,
\alpha_{n-1}^s+\alpha_{n-1}^{c_s}
\right).
\label{eq:app_cs_LO_recovery}
\end{equation}
The scalar arguments are absent when $n=0$. Thus, the same Bell-polynomial coefficients determine the LO hierarchy, now with $K_j=B_j(-\alpha_0^s-\alpha_0^{c_s},\ldots,-\alpha_{j-1}^s-\alpha_{j-1}^{c_s})$.

Defining $x=r/(8c_s)$, the first three relations become
\begin{equation}
\begin{split}
F_{0,0}^{(c_s)}&=-x,\\
F_{1,0}^{(c_s)}&=x\left(\alpha_0^s+\alpha_0^{c_s}+x\right),\\
F_{2,0}^{(c_s)}&=x\left[\alpha_1^s+\alpha_1^{c_s}-(\alpha_0^s+\alpha_0^{c_s})^2-3x(\alpha_0^s+\alpha_0^{c_s})-2x^2\right].
\end{split}
\end{equation}
In particular, the first relation gives $n_t=-r/(8c_s)$, while the higher relations include the evolution of the sound speed through $\alpha_j^{c_s}$. Both implementations reproduce these expressions through $n=8$, and substituting the LO observables recovers the background result $-2D_N^n\epsilon_1$. Setting $c_s=1$ and $\alpha_j^{c_s}=0$ reduces the entire hierarchy to the canonical expressions.

\subsection{NLO Results}

\subsubsection{Canonical Inflation}

We next consider the first corrections to the LO consistency relations.  The primordial spectra derived in Appendix~\ref{app:spectra_proof} give $r=16\epsilon_1(1+C\epsilon_2)+\mathcal O(\mathrm{SR}^3)$ and $\alpha_0^t=-2\epsilon_1[1+\epsilon_1+(C+1)\epsilon_2]+\mathcal O(\mathrm{SR}^3)$.  Here and below, $C\equiv\gamma_{\mathrm E}+\ln2-2$, where $\gamma_{\mathrm E}$ is the Euler--Mascheroni constant.  At this order, the background variable $\epsilon_2$ can be eliminated using $\epsilon_2=-\alpha_0^s-r/8+\mathcal O(\mathrm{SR}^2)$. Substituting this relation and expressing $\epsilon_1$ in terms of $r$, the constant $C$ cancels, leaving
\begin{equation}
F_{0,1}=-\frac{r}{8}\left(1-\alpha_0^s-\frac{r}{16}\right).
\label{eq:app_NLO_tilt}
\end{equation}
This reproduces the standard NLO consistency relation obtained from the spectra of Ref.~\cite{Stewart:1993bc}.

\subsubsection{$k$-Inflation}

For $k$-inflation, the NLO calculation must also include the sound-speed dependence of the scalar source and the translation between the scalar and tensor reference scales. Expanding Eq.~\eqref{eq:cs_source} gives $g_s=3\epsilon_1+3(q_1+s_1)/2+\mathcal O(\mathrm{SR}^2)$. Together with $\chi_{1;0}=-2C/3$ and $h_s=1+\epsilon_1+s_1+\mathcal O(\mathrm{SR}^2)$, this determines the first-order scalar correction at its acoustic reference scale.

The scalar spectrum must then be translated to the common tensor scale. Writing $\ell\equiv\ln c_s$, this translation contributes $-(2\epsilon_1+q_1+s_1)\ell$ at first order. Combining the acoustic correction with the scale translation, we obtain
\begin{equation}
\begin{split}
\ln\frac{8\pi^2M_{\rm Pl}^2\epsilon_1c_s\mathcal P_s(k_t)}{H^2}
&=-2(C+1)\epsilon_1-Cq_1-(C+2)s_1-(2\epsilon_1+q_1+s_1)\ell+\mathcal O(\mathrm{SR}^2),\\
r&=16\epsilon_1c_s\left[1+Cq_1+(C+2)s_1+(2\epsilon_1+q_1+s_1)\ell\right]+\mathcal O(\mathrm{SR}^3).
\end{split}
\label{eq:app_cs_NLO_spectra}
\end{equation}

The tensor tilt maintains its canonical background expression. We can therefore eliminate $\epsilon_1$ using the corrected tensor-to-scalar ratio and replace the remaining background derivatives by their LO reconstructions, $q_1=-2\epsilon_1-\alpha_0^s-\alpha_0^{c_s}$ and $s_1=\alpha_0^{c_s}$. This gives
\begin{equation}
F_{0,1}^{(c_s)}=-x\left[1-\frac{x}{2}-(1-\ell)\alpha_0^s-(C+3)\alpha_0^{c_s}\right].
\label{eq:app_cs_NLO_tilt}
\end{equation}
The scalar amplitude and the resulting consistency relation have also been compared with independently constructed analytic expressions. In the canonical limit, $\ell=\alpha_0^{c_s}=0$, Eq.~\eqref{eq:app_cs_NLO_tilt} reduces to Eq.~\eqref{eq:app_NLO_tilt}.

\subsection{N2LO Results}

\subsubsection{Canonical Inflation}

At N2LO, the calculation includes the quadratic corrections to both primordial spectra. Taking the corresponding spectral derivatives and eliminating the inflationary background through absolute SR order three yields
\begin{equation}
\begin{split}
F_{0,2}=F_{0,1}+\frac{r}{8}\Bigg[
\alpha_1^s-\frac{\pi^2}{12}(\alpha_0^s)^2
-\left(1+\frac{\pi^2}{6}\right)\frac{r}{8}\alpha_0^s
-\left(\frac34+\frac{\pi^2}{12}\right)\left(\frac{r}{8}\right)^2
\Bigg].
\end{split}
\label{eq:app_N2LO_tilt}
\end{equation}
This reproduces the N2LO consistency relation obtained from the independently derived second-order spectra of Refs.~\cite{Gong:2001he,Leach:2002ar}, after adopting the same reference convention. The comparison is performed coefficient by coefficient in the observable expansion, rather than for selected background parameters.

\subsubsection{$k$-Inflation}

At this order, the sound-speed dependence of the scalar amplitude provides an additional check of the construction. Following the exact mapping between scalar and tensor mode equations developed in Ref.~\cite{BeltranJimenez:2013ikr}, we can independently calculate the scalar coefficients by treating the acoustic scalar equation as an effective tensor equation. Introducing $\widetilde a=a\sqrt{\epsilon_1/c_s}$ and $d\widetilde N=Q_s\,dN$, the effective clock and first flow parameter are
\begin{equation}
\widetilde h=Q_sh_s,\qquad
\widetilde\epsilon_1=\frac{1}{Q_s}\left[\epsilon_1+\frac{q_1+s_1}{2}-D_N\ln Q_s\right],
\qquad Q_s=1+\frac{q_1-s_1}{2}.
\label{eq:app_cs_effective_background}
\end{equation}
The normalized mode equation is identical in the acoustic scalar and effective tensor descriptions. Since the spectral corrections satisfy $M_A=\ln|J_A|^2-2\ln h_A$, their amplitudes are related by $M_s=\widetilde M_t+2\ln Q_s$ at the acoustic reference scale. Here $\widetilde M_t$ is the canonical tensor correction evaluated on the effective background, with $D_{\widetilde N}=Q_s^{-1}D_N$.

This provides an independent way to calculate the scalar sound-speed contributions using the known tensor coefficients, rather than the direct acoustic-source expansion. To isolate these contributions, we denote by $\Delta M_s^{(j)}$ the terms of exactly SR order $j$ that disappear when all $s_i$ are set to zero. At quadratic order, we find
\begin{equation}
\begin{split}
\Delta M_s^{(2)}={}&
\left[\frac{\pi^2}{2}-3(C+2)\right]\epsilon_1s_1
+\left[\frac{\pi^2}{4}-(C+2)\right]q_1s_1\\
&+\left[\frac{\pi^2}{8}-(C+2)\right]s_1^2
+\left[\frac{\pi^2}{24}-\frac{(C+2)^2}{2}\right]s_2.
\end{split}
\label{eq:app_cs_N2LO_amplitude}
\end{equation}
These coefficients agree with those obtained directly from the acoustic scalar source. Since the tensor spectrum remains unchanged, we can then translate the scalar spectrum to the tensor reference scale and eliminate the background through absolute order three. The resulting consistency relation is
\begin{equation}
\begin{split}
F_{0,2}^{(c_s)}=F_{0,1}^{(c_s)}+x\Bigg[&
\left(1-\ell+\frac{\ell^2}{2}\right)\alpha_1^s
-\left(\frac{\pi^2}{12}-\ell+\frac{\ell^2}{2}\right)(\alpha_0^s)^2\\
&-\left(1+\frac{\pi^2}{6}-\ell\right)x\alpha_0^s
-\left(\frac34+\frac{\pi^2}{12}\right)x^2\\
&-\left(C+3+\frac{\pi^2}{6}\right)x\alpha_0^{c_s}
+\left[(C+3)(\ell-2)+\frac{\pi^2}{12}\right]\alpha_0^s\alpha_0^{c_s}\\
&-\left[\frac{(C+3)^2-1}{2}-\frac{\pi^2}{24}\right](\alpha_0^{c_s})^2
+\left[\frac{(C+3)^2+1}{2}-\frac{\pi^2}{24}\right]\alpha_1^{c_s}\Bigg].
\end{split}
\label{eq:app_cs_N2LO_tilt}
\end{equation}
Both implementations reproduce this expression. Moreover, the complete relations for $n=0,\ldots,3$ satisfy the forward-substitution identity in Eq.~\eqref{eq:app_validation_forward}, while setting $c_s=1$ and all sound-speed derivatives to zero recovers Eq.~\eqref{eq:app_N2LO_tilt}.

\subsection{N3LO Results}

\subsubsection{Canonical Inflation}

We now include the cubic corrections to the primordial spectra and eliminate the background through absolute SR order four. For the tensor tilt, the resulting relation is
\begin{equation}
\begin{split}
F_{0,3}=F_{0,2}+\frac{r}{8}\Bigg[&
\alpha_2^s
+\left(\frac{\pi^2}{12}-\frac{\zeta(3)}4\right)(\alpha_0^s)^3
+\left(1-\frac{\pi^2}{6}-\zeta(3)\right)\alpha_0^s\alpha_1^s\\
&+\left(\frac12+\frac{3\pi^2}{8}+2\zeta(3)\right)\frac{r}{8}(\alpha_0^s)^2
-\left(1+\frac{\pi^2}{6}+\zeta(3)\right)\frac{r}{8}\alpha_1^s\\
&+\left(\frac{11}{4}+\frac{5\pi^2}{12}+\frac{19\zeta(3)}4\right)
\left(\frac{r}{8}\right)^2\alpha_0^s
+\left(\frac{15}{8}+\frac{\pi^2}{8}+\frac{5\zeta(3)}2\right)
\left(\frac{r}{8}\right)^3\Bigg].
\end{split}
\label{eq:app_N3LO_tilt}
\end{equation}
This reproduces the N3LO result independently derived from the cubic spectra in Eqs.~(44) and~(54) of Ref.~\cite{Auclair:2022yxs}, with coefficient-by-coefficient agreement within a numerical tolerance of $10^{-9}$. We have extended the comparison to $n=1,2$ at each order $p=0,1,2,3$, finding agreement with the corresponding analytic results to the same accuracy.

\subsubsection{$k$-Inflation}

For $k$-inflation, we again use the effective-background construction as an independent check of the scalar coefficients. At cubic order, the canonical tensor spectrum is known through N3LO, allowing us to evaluate $\widetilde M_t+2\ln Q_s$ and extract the terms involving the sound-speed hierarchy. We obtain
\begin{equation}
\begin{split}
\Delta M_s^{(3)}={}&
\left[-5(C+2)+2\pi^2-7\zeta(3)\right]\epsilon_1^2s_1\\
&+\left[-\frac52(C+2)^2+\left(\frac{\pi^2}{2}-1\right)(C+2)+\frac{29\pi^2}{24}-7\zeta(3)\right]\epsilon_1q_1s_1\\
&+\left[-4(C+2)+\frac{5\pi^2}{4}-\frac{7\zeta(3)}2\right]\epsilon_1s_1^2
+\left[-2(C+2)^2+\left(\frac{\pi^2}{2}-2\right)(C+2)+\frac{\pi^2}{6}\right]\epsilon_1s_2\\
&+\frac{\pi^2-7\zeta(3)}{4}q_1^2s_1
+\left[-(C+2)+\frac{\pi^2}{2}-\frac{7\zeta(3)}4\right]q_1s_1^2\\
&+\left[-\frac{(C+2)^2}{2}+\left(\frac{\pi^2}{4}-1\right)(C+2)+\frac{\pi^2}{24}\right]q_1s_2\\
&+\left[-(C+2)^2+\left(\frac{\pi^2}{4}+2\right)(C+2)-\frac{5\pi^2}{12}\right]q_2s_1\\
&+\left[-(C+2)+\frac{\pi^2}{4}-\frac{7\zeta(3)}{12}\right]s_1^3
+\left[-\frac32(C+2)^2+\left(\frac{\pi^2}{4}-1\right)(C+2)+\frac{\pi^2}{8}\right]s_1s_2\\
&+\left[-\frac{(C+2)^3}{6}+\frac{\pi^2(C+2)}{24}-\frac{\zeta(3)}3\right]s_3.
\end{split}
\label{eq:app_cs_N3LO_amplitude}
\end{equation}

Together with the canonical scalar contributions, these terms reproduce the complete acoustic log amplitude calculated from the universal source coefficients. The comparison is performed before translating the scalar spectrum to the tensor reference scale, so it provides a direct check of the acoustic calculation. In particular, it verifies the contributions involving $s_2$ and $s_3$, together with the mixed Hubble and sound-speed terms that cannot be isolated by considering constant $c_s$.

We further compare our results with the independent analytic N3LO spectra of Ref.~\cite{Bianchi:2024qyp}. For a direct comparison, we specialize the general expressions of that reference to $Z=2M_{\rm Pl}^2\epsilon_1/c_s^2$ and propagation speed $c_s$ in the scalar sector, and to constant $Z$ and unit propagation speed in the tensor sector. We then express the corresponding flow parameters in terms of our background hierarchy and adopt the same reference-scale conventions, $-k\tau=1$ for scalars and $-k\eta=1$ for tensors. With these identifications, we find exact agreement with Ref.~\cite{Bianchi:2024qyp} for the analytic coefficients of all $31$ scalar and $8$ tensor monomials through N3LO, including the leading-order contributions. This agreement is obtained with both the iterative and closed implementations. Repeating the comparison with the independently determined numerical coefficients, we find a maximum absolute difference below $4\times10^{-14}$. This provides an independent validation of the primordial spectra at their respective reference scales, before the translation to a common tensor reference scale and the subsequent elimination of the inflationary background.

After applying the reference-scale translation in Eq.~\eqref{eq:cs_pivot_spectrum}, both implementations give identical consistency relations for $n=0,\ldots,3$ at $p=3$. Each relation satisfies the forward-substitution identity in Eq.~\eqref{eq:app_validation_forward} for the generated coefficient values. The canonical limit of the tensor tilt reproduces Eq.~\eqref{eq:app_N3LO_tilt}. The analytic amplitude coefficients displayed above and the resulting tensor-tilt coefficients agree with the numerical calculation.

\subsection{N10LO Results}

\subsubsection{Canonical Inflation}

Finally, we test the construction at substantially higher SR order. These checks are not intended to reproduce a known general N10LO consistency relation, but to verify that the same procedure continues to give consistent results when the number of background derivatives and source coefficients becomes much larger.

We first consider the exactly solvable limit of constant $\epsilon_1=\epsilon$, with all higher flow parameters vanishing. In this case, the scalar and tensor mode equations admit Hankel solutions, and their logarithmic corrections are identical:
\begin{equation}
M_s=M_t=(2\nu-3)\ln2
+2\ln\frac{\Gamma(\nu)}{\Gamma(3/2)}+2\ln(1-\epsilon),
\qquad
\nu=\frac{3-\epsilon}{2(1-\epsilon)}.
\end{equation}
The SR expansion generated by our construction reproduces the Taylor series of this exact expression through tenth order. The corresponding observables satisfy $r=16\epsilon$ and $\alpha_0^s=\alpha_0^t=-2\epsilon/(1-\epsilon)$, with all higher runnings equal to zero.

We then examine a nonconstant background, for which the mode equations are integrated independently. The scalar and tensor amplitudes are compared through tenth order using $a(x)=x^{-1}\exp[-e(x^b-1)/b]$. To isolate the contributions at successive SR orders, we rescale the parameters according to $(e,b)\rightarrow(\lambda e,\lambda b)$ and compare the coefficients of the resulting expansion. The initial and final integration times are refined independently to control numerical errors. In addition, all $138$ universal source coefficients needed at this order are checked using the separate mode and growing-pump solver described above.

Having verified the spectral ingredients, we turn to the consistency relations. We consider the tensor running, $n=1$, through relative order $p=10$, which requires absolute SR order $W=12$. Substituting the corresponding forward observable map into the generated expression gives
\begin{equation}
\left[
F_{1,10}\big(\mathcal U_{12}(\mathbf b)\big)
-\alpha_1^t(\mathbf b)
\right]_{\leq12}=0.
\end{equation}
This identity holds exactly for the generated coefficient values. Every term has the required SR weight, the final relation depends only on observables, and it vanishes when $r=0$. The forward substitution therefore provides a direct test of the high-order background elimination, complementary to the independent mode integrations used to validate the universal coefficients.

\subsubsection{$k$-Inflation}

The high-order sound-speed calculation can also be tested against exact solutions. We consider backgrounds with constant $\epsilon_1=\epsilon$ and $s_1=s$, corresponding to $c_s\propto e^{sN}$, while $q_j=0$ and $s_j=0$ for $j\geq2$. Both scalar and tensor mode equations are then solved by Hankel functions, with
\begin{equation}
h=\frac{1}{1-\epsilon},\qquad
h_s=\frac{1}{1-\epsilon-s},\qquad
\nu_t=\frac12+h,\qquad
\nu_s=\frac12+h_s\left(1-\frac{s}{2}\right).
\end{equation}
At their respective reference scales, the exact logarithmic amplitudes are
\begin{equation}
M_A=(2\nu_A-3)\ln2+2\ln\frac{\Gamma(\nu_A)}{\Gamma(3/2)}-2\ln h_A,
\qquad h_t=h.
\label{eq:app_cs_Hankel_amplitudes}
\end{equation}

An additional advantage of this family is that the scalar tilt is constant and known exactly: $3-2\nu_s=-(2\epsilon+s)/(1-\epsilon-s)$. The translation from the scalar reference scale to the tensor reference scale must therefore add $(3-2\nu_s)\ln(c_sh_s/h)$ to the scalar logarithmic amplitude. This provides an independent analytic check not only of the acoustic mode normalization, but also of the reference-scale translation introduced in Appendix~\ref{app:noncanonical}.

We compare the coefficients of the expansion under $(\epsilon,s)\rightarrow(\lambda\epsilon,\lambda s)$ through $\lambda^{10}$, using $s/\epsilon=-0.4,0.3,0.8$ and $c_{s*}=0.5,0.8,1.2$. The largest difference between the calculated and exact coefficients, normalized by one plus the absolute reference coefficient, is below $5\times10^{-15}$.

To test a less restricted sound-speed evolution, we also compare both implementations on the background family
\begin{equation}
q_j=(j-1)!\,q_1^j,\qquad
s_j=(j-1)!\,q_1^{j-1}s_1.
\end{equation}
For this family, the clocks, sources, and logarithmic amplitudes agree exactly through tenth order. The effective-background construction in Eq.~\eqref{eq:app_cs_effective_background} independently reproduces the same scalar result. We also verify the conversion to the spectral sound-speed hierarchy and the recovery of the canonical limit at the same order.

These N10LO comparisons verify the primordial spectra against exact and independently integrated solutions for the background families considered above. The consistency relations are also tested without restricting the inflationary background to these particular examples. In particular, all $32$ published $k$-inflation relations satisfy the exact forward-substitution identity, including the tensor tilt through N10LO. The complete agreement between the iterative and closed background inversions has been established separately through N3LO. Together, these checks test the spectral construction at high SR order and the consistency of the general observable inversion.

\clearpage
\twocolumngrid
\bibliography{main}

\end{document}